\documentclass{IEEEtaes}
\usepackage{amsmath,amsfonts,amssymb}
\usepackage{algorithm}
\usepackage{algpseudocode}
\usepackage{array}
\usepackage[caption=false,font=normalsize,labelfont=sf,textfont=sf]{subfig}
\usepackage{textcomp}
\usepackage{stfloats}
\usepackage{url}
\usepackage{verbatim}
\usepackage{graphicx}
\usepackage{cite}
\usepackage{color}
\usepackage{booktabs}
\usepackage{multirow}
\usepackage{siunitx}
\usepackage{makecell}

\algrenewcommand\algorithmicrequire{\textbf{Input:}}
\algrenewcommand\algorithmicensure{\textbf{Output:}}

\newtheorem{proposition}{Proposition}

\newtheorem{remark}{Remark}
\newtheorem{example}{Example}

\allowdisplaybreaks

\begin{document}

\title{{Efficient Implementations of Extended Object PMBM Filters with Blocked Gibbs Sampling}}

\author{Yuxuan Xia}
\member{Member, IEEE}

\author{{\'A}ngel F. Garc{\'\i}a-Fern{\'a}ndez}

\author{Lennart Svensson}
\member{Senior Member, IEEE}



\authoraddress{Y. Xia is with the School of Automation and Intelligent Sensing, Shanghai Jiao Tong University, 200240 Shanghai, China
(e-mail: \href{mailto:yuxuan.xia@sjtu.edu.cn}{yuxuan.xia@sjtu.edu.cn}). Á. F. García-Fernández is with the Information Processing and Telecommunications Center, ETSI de Telecomunicación, Universidad Politécnica de Madrid, 28040 Madrid, Spain. L. Svensson is with the Department of Electrical Engineering, Chalmers University of Technology, 41296 Gothenburg, Sweden.\\
This work was supported by the National Natural Science Foundation of
China under grant No. 62501384.}

\maketitle

\begin{abstract}
This paper considers multiple extended object tracking based on Poisson multi-Bernoulli mixture (PMBM) filtering, which gives the closed-form Bayesian solution for standard multiple extended object models with Poisson birth. To efficiently address the challenging extended object data association problem in PMBM filtering, we develop implementations of the extended object PMBM filter using blocked Gibbs sampling. By formulating the PMBM density on an augmented state space with auxiliary variables and leveraging the Poisson object measurement model, we first derive a joint posterior over potential objects, previous global hypotheses, and current measurement association variables, together with its corresponding factorization. This factorized representation leads to blocked Gibbs samplers that can efficiently generate high-weight global hypotheses and thereby provide an efficient implementation of the PMBM update step. We further introduce a collapsed Gibbs sampling variant, in which the Bernoulli object existence variables are marginalized out, yielding higher sampling efficiency, especially for the initiation of newly detected objects. The proposed methods, implemented under the gamma Gaussian inverse-Wishart model, are compared with existing extended object PMBM baselines with different data association strategies. Simulation results demonstrate that the proposed approaches achieve comparable tracking performance while requiring substantially less runtime.
\end{abstract}


\section{Introduction}

Multi-object tracking (MOT) considers the recursive estimation of multi-object states from noisy sensor measurements acquired over time \cite{bar2011tracking,meyer2018message,streit2021analytic}. It underpins a broad set of applications in, e.g., defense \cite{zhu2021detection}, surveillance \cite{cioppa2022soccernet}, and autonomous driving \cite{huang2025vehicle}. Conventional MOT methods often consider the point object assumption, under which each object contributes at most one measurement at each sensor scan. For modern high-resolution sensors such as Lidar and automotive radar, this assumption is frequently violated since a single object may generate multiple measurements in one sensor scan. Tracking objects under this measurement regime is commonly referred to as extended object tracking (EOT); see \cite{extendedoverview,granstrom2022tutorial} for overviews. In EOT, the object state typically includes both kinematic variables and an extent description that captures shape and size. In this paper, we consider the multiple EOT setting with an unknown and time-varying number of objects.

In EOT, the set of measurements generated by the object at a time step is typically modeled as an inhomogeneous Poisson point process (PPP) \cite{ppp}, where the number of measurements follows a Poisson distribution, and the individual measurements are conditionally independently and identically distributed given the object state. A key advantage of this formulation is that it does not require explicit associations between measurements and specific reflection points on the object, which makes it particularly convenient for EOT. The PPP measurement model has been widely adopted in multiple EOT methods, including, e.g., multiple hypothesis tracking (MHT) \cite{coraluppi2018multiple}, joint probabilistic data association (JPDA) filters \cite{vivone2016joint,yang2018linear}, graph-based approaches built on message passing \cite{meyer2020scalable,florian2021scalable,li2024multisensor,guo2025gaussian,ma2026closed,ma2024max}, and random finite set (RFS)-based filters \cite{phdextended2,cphdextended,lmbextended,pmbmextended2,xia2021poisson,xia2023trajectory}. 

A major difficulty in multiple EOT lies in data association, since the measurement-to-object correspondences are unknown. Making this step computationally tractable is essential for practical multiple EOT implementations. A widely used way to address it is a two-stage clustering-and-assignment (C\&A) procedure \cite{lmbextended,pmbmextended2}. In this approach, clustering algorithms, e.g., distance partitioning \cite{phdextended2} or DBSCAN \cite{ester1996density}, are first applied to find multiple plausible measurement partitions. Then, for each measurement partition, we typically obtain the $k$-best assignments by solving optimal 2D assignment problems and using Murty's algorithm \cite{crouse2016implementing}. C\&A-based implementations are usually effective when the objects are well separated, but their performance may seriously deteriorate once objects become closely spaced. In such difficult cases, sampling-based approaches, such as \cite{fatemi2017poisson,soextended}, can provide better estimation performance by operating directly on the data association likelihood and drawing association hypotheses without an explicit clustering stage, although this often comes at a considerable computational cost.

Both C\&A-based and sampling-based approaches rely on explicitly constructing global association hypotheses with non-negligible weights, while hypotheses with very small weights must be discarded to maintain tractability. A more scalable alternative is to work directly with the marginal multi-object posterior, in which the association uncertainty is integrated out, leading to a multi-Bernoulli (MB) approximation of the multi-object posterior. Existing work has mainly pursued this idea in two ways: 1) applying (loopy) belief propagation (BP) to a factor-graph representation of the joint posterior over data association variables and multi-object states, with cardinality uncertainty represented through either binary object existence variables or Bernoulli distributions \cite{meyer2020scalable,florian2021scalable,xia2023trajectory,li2024multisensor,guo2025gaussian}; or 2) adopting mean-field variational inference to approximate the joint posterior \cite{gan2024variational,cheng2025variational,yang2025three,gan2025pivot}.

Specifically, \cite{meyer2020scalable,florian2021scalable,xia2023trajectory} employ particle BP for multiple EOT. The simulation study in \cite{xia2023trajectory} further shows that, for the same underlying multi-object filter, a particle BP-based implementation outperforms both C\&A-based \cite{pmbmextended2} and sampling-based implementations \cite{soextended} in a difficult scenario, in which the objects initially move close to each other before separating. Analytical BP implementations have also been developed in \cite{guo2025gaussian} for Gaussian models and in \cite{li2024multisensor} for the gamma Gaussian inverse-Wishart (GGIW) models. However, these implementations involve merging object state densities corresponding to different association outcomes obtained using clustering, which can degrade estimation performance in challenging scenarios with closely-spaced objects. Instead of sum-product message passing, \cite{ma2024max} employs the max-sum message passing rule to infer the maximum a posteriori (MAP) configurations of data association variables, together with a light-weight GGIW implementation. However, retaining only the MAP estimate can result in degraded performance in scenarios with significant data association ambiguity.

In \cite{ma2026closed}, the BP is used for computing the marginal data association probabilities while the mean-field variational inference is employed for analytically updating the multi-object state densities. For tracking scenarios with a fixed and known number of objects, the method proposed in \cite{ma2026closed} is reported to outperform existing BP-based methods. Mean-field variational approximations have been used for joint inference of multi-object state and data association in \cite{gan2024variational,cheng2025variational,yang2025three}, but \cite{gan2024variational,cheng2025variational} assume a fixed and known number of objects, whereas \cite{yang2025three} handles object birth and death via heuristic mechanisms. The approach in \cite{gan2024variational} is further extended in \cite{gan2025pivot} to estimate the object existences by employing a two-stage variational inference.

While marginalizing data association uncertainty can yield efficient multiple EOT algorithms, retaining it in the full multi-object posterior captures the mixture density structure induced by multiple global association hypotheses and avoids premature hypothesis selection or merging under ambiguity. In \cite{li2023adaptive}, a scalable Rao-Blackwellized sequential Gibbs sampling scheme is presented for multiple EOT with PPP measurement model, targeting the joint posterior of measurement association variables and object states. 
The PPP measurement model induces conditional independence among measurement association variables given the multi-object state, enabling parallel sampling and efficient exploration of data association hypotheses. This is an important advantage over previous sampling-based methods used in \cite{fatemi2017poisson,soextended}, where the measurement associations of at most two potential objects (Bernoulli components) can be modified at each sampling iteration. Nevertheless, the formulation in \cite{li2023adaptive} assumes a fixed and known number of objects. It has been later extended in \cite{li2023scalable} to multiple EOT, but under the restrictive assumption of a single Bernoulli birth and the sampling process is no longer fully parallelizable.

For the standard multi-object model with a PPP birth and the object measurement model, in which each object independently generates a set of measurements, the exact multi-object posterior takes the Poisson multi-Bernoulli mixture (PMBM) form \cite{garcia2021poisson,garcia2023poisson}. An important feature of the PMBM representation is that it compactly encodes association hypotheses while capturing probabilistic object existence within individual Bernoulli components. It also supports measurement-driven track initiation through a PPP representation of undetected objects \cite{pmbmpoint2}. PMBM filtering recursions have been developed for point objects \cite{pmbmpoint2}, extended objects \cite{pmbmextended2}, and more general measurement models \cite{garcia2021poisson,garcia2023poisson}. In addition, the same framework can preserve full trajectory information by working with sets of trajectories \cite{granstrom2024poisson,xia2023trajectory,xia2023efficient}. The PMBM recursion also accommodates an MB birth model. Specifically, the multi-Bernoulli mixture (MBM) filtering recursion can be achieved
by setting the Poisson intensity for undetected objects to zero and introducing new Bernoulli components in the prediction step \cite{pmbmpoint2,garcia2019gaussian}.

The extended object PMBM filter and its approximation, the extended object Poisson multi-Bernoulli (PMB) filter \cite{xia2021poisson}, have been realized with both C\&A-based \cite{pmbmextended2} and sampling-based implementations \cite{soextended}. In addition, a particle BP implementation of the extended object PMB filter is presented in \cite{xia2023trajectory}. Under the same underlying data association strategy, these PMBM and PMB filters have been shown to deliver competitive performance relative to other multi-object filters. Extended object PMBM filters have also been used in a variety of applications, including MOT with Lidar \cite{soextended,ding2024lidar,baerveldt2024multiple}, 4D radar \cite{liu2024framework}, roadside radar-camera sensing \cite{deng20243}, lane-marking detections \cite{xia2024bayesian}, mapping with radio-frequency 5G signals \cite{ge20205g}, and joint MOT with occupancy-grid mapping \cite{baerveldt2026combining}.

Motivated by the advantages of PMBM filtering \cite{pmbmextended2,xia2021poisson} and by the scalable sampling-based association strategy for the PPP measurement model in \cite{li2023adaptive}, this paper proposes efficient  implementations of the extended object PMBM filter with blocked Gibbs sampling. To this end, we formulate the PMBM density on an augmented state space with auxiliary variables and derive a joint posterior over set of potential objects, previous global hypotheses, and current measurement association variables, together with its corresponding factorization. A related factorized joint posterior, without the previous global hypotheses, was used in \cite{xia2023trajectory} to develop a BP-based implementation. Here we instead use the factorized joint posterior to derive the conditional distributions required for blocked Gibbs sampling. An extended object trajectory PMB filter with blocked Gibbs sampling was presented in a preliminary work, which preserves the full trajectory information and performs MB approximation using sample statistics \cite{xia2023efficient}. The resulting implementation was later applied to multi-lane tracking and validated on real data in \cite{xie2026bfmap}.

In this paper, we focus on efficient implementations of the extended object PMBM update via blocked Gibbs sampling and the estimation of current object states. The main contributions are:
\begin{itemize}
  \item We derive a factorized joint posterior representation of the previous global hypothesis, current measurement associations, and set of potential objects. The factorization is achieved by representing the PMBM posterior on an augmented space with auxiliary variables and leveraging the PPP measurement model.
  \item We develop an implementation of the extended object PMBM update step via blocked Gibbs sampling, which enables the efficient generation of high-weight global hypotheses. In each  iteration, the three considered blocks of variables are sequentially sampled from their respective conditional densities under the factorized joint posterior.
  \item We also develop an implementation of the extended object PMBM update step via collapsed Gibbs sampling, in which the Bernoulli object existence variables are marginalized out from sampling. Although this introduces approximations, it improves sampling efficiency, especially for the initiation of newly detected objects.
  \item We introduce track-oriented implementations of the extended object PMBM filters that avoid redundant computations across sampling iterations. Moreover, we adopt a weight-based approximation of the multi-object posterior. The blocked Gibbs sampler is used to find global hypotheses, whose weights are computed using the PMBM update equations rather than their empirical sampling frequencies. Consequently, unique global hypotheses found during all sampling iterations contribute to the posterior approximation.
  \item We compare the proposed methods under the GGIW model with two extended object PMBM implementations using C\&A \cite{pmbmextended2} and stochastic optimization \cite{soextended} as well as an extended object PMB filter using particle BP \cite{xia2023trajectory}. Simulation results demonstrate that the proposed methods achieve comparable estimation performance with substantially less runtime.
\end{itemize}

In the rest of the paper, Section~II reviews the multi-object model and extended object PMBM filtering. Section III presents the joint posterior and its factorization. In Section IV, we present the extended object PMBM filter based on blocked Gibbs sampling. Section V presents the GGIW implementation, simulation setup, and numerical results. Finally, Section VI concludes the paper.

\section{Background}

This section first reviews the standard multi-object dynamic model \cite{rfs} and the measurement model with Poisson object measurements. It then introduces the PMBM density together with its hypothesis structure, followed by the extended object PMBM filtering recursion.

For a generic space $D$, we denote $\mathcal{F}(D)$ for the collection of all finite subsets of $D$, and for any $A \in \mathcal{F}(D)$, its cardinality is denoted by $|A|$ \cite{rfs}. The symbol $\uplus$ denotes union over mutually disjoint sets, and we write $\langle f,g \rangle$ the inner product $\int f(x)g(x)\,dx$. Given a real-valued function $f$, we use the multi-object exponential representation $f^A$ to denote $\prod_{x\in A} f(x)$, with the convention that $f^\emptyset = 1$. Lastly, the Kronecker delta centered at $x$ is denoted $\delta_x[\cdot]$.

\subsection{Multi-Object Models}
\label{sec_model}

At time step $k$, the state of a single object $x_k$ is defined in a locally compact, Hausdorff, and second-countable space $\mathcal{X}$ \cite{rfs}. The single object state may include quantities of interest such as the kinematic state and the extent state, where the latter characterizes the object shape and size. The multi-object state at time step $k$ is written as $\mathbf{x}_k = \{x_k^1,\dots,x_k^{n_k}\} \in \mathcal{F}(\mathcal{X})$, whereas the corresponding measurement set is $\mathbf{z}_k = \{z_k^1,\dots,z_k^{m_k}\} \in \mathcal{F}(\mathbb{R}^{n_z})$. 

\subsubsection{Dynamic model}
At time step $k$, the set of newborn objects is a PPP with birth intensity $\lambda_k^B(\cdot)$. Each existing object $x_{k-1}$ at time step $k-1$ survives at time step $k$ with a probability $p_k^S(x_{k-1})$; conditioned on survival, it independently evolves according to the transition density $g_k(\cdot|x_{k-1})$.

\subsubsection{Measurement model}
The set of measurements $\mathbf{z}_k$ at time step $k$ is composed of object-originated measurements together with clutter. The clutter is a PPP with intensity $\lambda_k^C(z_k)=\gamma_k^C\mu_k^C(z_k)$, where $\gamma_k^C$ and $\mu_k^C(\cdot)$ are the clutter rate and spatial density, respectively.

Let $\mathbf{w}_k$ denote the set of measurements produced by object $x_k$. Following \cite{xia2023trajectory,xia2023efficient}, we model $\mathbf{w}_k$ as a PPP, whose set density is
\begin{equation}\label{eq_ppp_meas}
    \ell_k(\mathbf{w}_k|x_k) = e^{-\gamma_k(x_k)}\prod_{z_k \in \mathbf{w}_k}\gamma_k(x_k)\ell_k(z_k|x_k),
\end{equation}
where $\gamma_k(\cdot)$ is the Poisson rate and $\ell_k(\cdot|x_k)$ is the single measurement likelihood. This model corresponds to setting the explicit object detection probability to one in the more flexible zero-inflated Poisson measurement model \cite{pmbmextended2}. Nevertheless, an object produces no measurements with probability $e^{-\gamma_k(x_k)}$, and hence the effective probability of obtaining at least one measurement (effective probability of detection) is $1-e^{-\gamma_k(x_k)}$ \cite[Eq. (3)]{phdextended2}.

\subsection{PMBM Density}
Given the measurement sequence $\mathbf{z}_{1:k^\prime}$ up to time step $k^\prime \in \{k-1,k\}$ and the multi-object models in Section~\ref{sec_model}, the multi-object density at time step $k$ is a PMBM \cite{pmbmextended2}
\begin{align}
    f_{k|k^\prime}(\mathbf{x}_k) &= \sum_{\mathbf{x}^u_k \uplus \mathbf{x}^d_k = \mathbf{x}_k}f_{k|k^\prime}^p\left(\mathbf{x}_k^u\right)f_{k|k^\prime}^{mbm}\left(\mathbf{x}^d_k\right),\label{eq_pmbm}\\
    f_{k|k^\prime}^p\left(\mathbf{x}_k^u\right) &= e^{-\left\langle \lambda_{k|k^\prime},1\right\rangle}\left[ \lambda_{k|k^\prime}(\cdot) \right]^{\mathbf{x}_k^u},\label{eq_ppp}\\
    f_{k|k^\prime}^{mbm}\left(\mathbf{x}^d_k\right) &= \sum_{a\in\mathcal{A}_{k|k^\prime}}w^a_{k|k^\prime}\sum_{\uplus_{l=1}^{n_{k|k^\prime}}\mathbf{x}_k^l = \mathbf{x}_k^d}\prod_{i=1}^{n_{k|k^\prime}}f^{i,a^i}_{k|k^\prime}\left(\mathbf{x}_k^i\right),\label{eq_mbm}
\end{align}
where $f_{k|k-1}(\cdot)$ and $f_{k|k}(\cdot)$ denote the PMBM predicted density and the PMBM filtering density at time step $k$, respectively. In \eqref{eq_ppp}, $\lambda_{k|k^\prime}(\cdot)$ denotes the Poisson intensity for the undetected objects, which are objects hypothesized to exist but have never been detected. The MBM term in \eqref{eq_mbm} describes the set of potential objects, which have been detected at least once by time step $k^\prime$.

The MBM term in \eqref{eq_mbm} has $n_{k|k^\prime}$ Bernoulli components, and Bernoulli component $i$ has $h^i_{k|k^\prime}$ local hypotheses. For a local hypothesis $a^i \in \{1,\dots,h^i_{k|k^\prime}\}$, its corresponding Bernoulli density is given by
\begin{equation}
    f^{i,a^i}_{k|k^\prime}\left(\mathbf{x}_k^i\right) = \begin{cases}
        1 - r_{k|k^\prime}^{i,a^i} & \mathbf{x}_k^i = \emptyset \\
        r_{k|k^\prime}^{i,a^i} f^{i,a^i}_{k|k^\prime}(x) & \mathbf{x}_k^i = \{x\} \\
        0 & \text{otherwise},
    \end{cases}
\end{equation}
where $r^{i,a^i}_{k|k^\prime}$ denotes the existence probability and $f^{i,a^i}_{k|k^\prime}(\cdot)$ is the existence conditioned single-object density. The set $\mathcal{A}_{k|k^\prime}$ contains all global hypotheses \cite{xia2023trajectory}, and for each $a=(a^1,\dots,a^{n_{k|k^\prime}})\in\mathcal{A}_{k|k^\prime}$, the corresponding MB in \eqref{eq_mbm} gives the density of the set of detected objects conditioned on selecting one local hypothesis $a^i$ from each Bernoulli component. The weight of global hypothesis $a$ satisfies
\begin{equation}
  \textstyle 
    w^a_{k|k^\prime} \propto \prod_{i=1}^{n_{k|k^\prime}}w^{i,a^i}_{k|k^\prime},
\end{equation}
where $w^{i,a^i}_{k|k^\prime}$ is the weight associated with local hypothesis $a^i$ of Bernoulli component $i$, and the weights $w^a_{k|k^\prime}$ sum up to one.

We index the $j$-th measurement $z_k^j$ at time step $k$ by the index pair $(k,j)$, and denote by $\mathcal{M}_k$ the set of all such index pairs up to and including time step $k$. For the multi-object measurement model given in Section~\ref{sec_model}, the set $\mathcal{A}_{k|k^\prime}$ of global hypotheses is
\begin{align}
    \mathcal{A}_{k|k^\prime} = \Bigg\{ a = &\left(a^1,\dots,a^{n_{k|k^\prime}}\right): a^i \in \left\{1,\dots,h^i_{k|k^\prime}\right\}~\forall~i, \nonumber \\  & \biguplus_{i=1}^{n_{k|k^\prime}} \mathcal{M}_{k^\prime}^{i,a^i} = \mathcal{M}_{k^\prime} \Bigg\},
\end{align}
where $\mathcal{M}_{k^\prime}^{i,a^i}$ denotes the set of index pairs assigned to local hypothesis $a^i$ of Bernoulli component $i$.

\subsection{Extended Object PMBM Filtering}

Next, we review the PMBM prediction and update equations for the multi-object models in Section~\ref{sec_model} \cite{pmbmextended2}. For the update step, we adopt the hypothesis structure, where each measurement creates a new Bernoulli component \cite{xia2021poisson}, rather than introducing a new one for every non-empty measurement subset as in \cite{pmbmextended2}.

\begin{proposition}
Suppose the PMBM filtering density at time step $k-1$ has the form \eqref{eq_pmbm}. Then the predicted density is a PMBM of the form \eqref{eq_pmbm}, with $n_{k|k-1}=n_{k-1|k-1}$, $h^i_{k|k-1}=h^i_{k-1|k-1}$, $w^{i,a^i}_{k|k-1}=w^{i,a^i}_{k-1|k-1}$, and
    \begin{subequations}
        \begin{align}
            \lambda_{k|k-1}(x) &= \lambda^B_k(x) + \left\langle \lambda_{k-1|k-1},g_k(x|\cdot)p^S_k(\cdot) \right\rangle,\\
            r^{i,a^i}_{k|k-1} &= r^{i,a^i}_{k-1|k-1}\left\langle f^{i,a^i}_{k-1|k-1},p^S_k \right\rangle,\\
            f^{i,a^i}_{k|k-1}(x) &= \frac{\left\langle f^{i,a^i}_{k-1|k-1},g_k(x|\cdot)p^S_k(\cdot) \right\rangle}{\left\langle f^{i,a^i}_{k-1|k-1},p^S_k \right\rangle}.
        \end{align}
    \end{subequations}
    \label{pmbm_prediction} 
\end{proposition}
The prediction recursion in Proposition~\ref{pmbm_prediction} coincides with the PMBM prediction in \cite{pmbmpoint,pmbmextended2} and does not depend on the specific measurement model.

\begin{proposition}
Suppose the predicted PMBM density has the form \eqref{eq_pmbm}, and we let ${\bf z}_k = \{z_k^1,\dots,z_k^{m_k}\}$. Then the updated density remains of PMBM form \eqref{eq_pmbm}, with
\begin{align}
  n_{k|k} &= n_{k|k-1} + m_k, \\
  \mathcal{M}_k = \mathcal{M}_{k-1} &\cup \left\{(k,j) \mid j \in \{1,\dots,m_k\} \right\},\\
  \lambda_{k|k}(x) &= \ell_k(\emptyset|x)\lambda_{k|k-1}(x).\label{eq_likelihood_missed_ppp}
\end{align}

For each Bernoulli component $i\in\{1,\dots,n_{k|k-1}\}$ in the predicted PMBM density, there are $h^i_{k|k}=2^{m_k}h^i_{k|k-1}$ updated local hypotheses, corresponding to a misdetection and updates with non-empty subsets of ${\bf z}_k$. For the misdetection case, with $a^i\in\{1,\dots,h^i_{k|k-1}\}$, we have $\mathcal{M}_k^{i,a^i}=\mathcal{M}_{k-1}^{i,a^i}$, and
\begin{subequations}
  \begin{align}
    \ell_{k|k}^{i,a^i,0} &= \left\langle f_{k|k-1}^{i,a^i},\ell_k(\emptyset|\cdot) \right\rangle,\label{eq_likelihood_missed}\\
    w_{k|k}^{i,a^i} &= w_{k|k-1}^{i,a^i}\left(1 - r_{k|k-1}^{i,a^i} + r_{k|k-1}^{i,a^i}\ell_{k|k}^{i,a^i,0}\right),\\
    r_{k|k}^{i,a^i} &= \frac{r_{k|k-1}^{i,a^i}\ell_{k|k}^{i,a^i,0}}{1 - r_{k|k-1}^{i,a^i} + r_{k|k-1}^{i,a^i}\ell_{k|k}^{i,a^i,0}},\\
    f_{k|k}^{i,a^i}(x) &= \frac{\ell_k(\emptyset|x)f_{k|k-1}^{i,a^i}(x)}{\ell_{k|k}^{i,a^i,0}}.
  \end{align}
\end{subequations}

We let ${\bf w}_k^1,\dots,{\bf w}_k^{2^{m_k}-1}$ denote the $2^{m_k}-1$ non-empty subsets of ${\bf z}_k$. Consider the $i$-th Bernoulli component with $i\in\{1,\dots,n_{k|k-1}\}$, and let $\tilde{a}^i\in\{1,\dots,h^i_{k|k-1}\}$ be one of its local hypotheses in the predicted PMBM density. For a measurement subset ${\bf w}_k^j$, where $j\in\{1,\dots,2^{m_k}-1\}$, the corresponding updated local hypothesis is assigned the index
$a^i=\tilde{a}^i+h^i_{k|k-1}j$,
and
\begin{subequations}
  \begin{align}
    \mathcal{M}_{k}^{i,a^i} &= \mathcal{M}_{k-1}^{i,\tilde{a}^i} \cup \left\{ (k,p): z_k^p \in {\bf w}_k^j \right\},\\
    \ell_{k|k}^{i,a^i,j} &= \left\langle f_{k|k-1}^{i,\tilde{a}^i},\ell_k\left({\bf w}_k^j|\cdot\right) \right\rangle, \label{eq_likelihood_Bernoulli_update}\\
    w_{k|k}^{i,a^i} &= w_{k|k-1}^{i,\tilde{a}^i}r_{k|k-1}^{i,\tilde{a}^i}\ell_{k|k}^{i,a^i,j},\\
    r_{k|k}^{i,a^i} &= 1,\\
    f_{k|k}^{i,a^i}(x) &= \frac{\ell_k\left({\bf w}_k^j|x\right)f_{k|k-1}^{i,\tilde{a}^i}(x)}{\ell_{k|k}^{i,a^i,j}}.
  \end{align}
\end{subequations}

Each newly created Bernoulli component is associated with its own collection of measurement subsets and has a different $h^i_{k|k}$. Following \cite{xia2021poisson}, for the $i$-th new Bernoulli component with $i\in\{1,\dots,m_k\}$, we let ${\cal S}_i$ be the set of associated measurement subsets, recursively defined as
\begin{equation}\label{eq_new_local_meas}
  {\cal S}_i = \left\{ \left\{ {z}_k^i \right\} \right\} \cup \left( \cup_{{\bf w}\in \cup_{j=1}^{i-1}{\cal S}_j} \left\{ \left\{ { z}_k^i \right\} \cup {\bf w}  \right\} \right),
\end{equation}
with ${\cal S}_1 = \{ \{ {z}_k^1 \} \}$ and $|\mathcal{S}_i|=2^{i-1}$. This construction ensures that the $i$-th new Bernoulli component must include the local hypothesis generated by the single measurement $z_k^i$, and it cannot involve measurements with index exceeding $i$. We further write ${\bf w}_k^{i,\iota}\in\mathcal{S}_i$ for the $\iota$-th non-empty subset of measurements associated with Bernoulli component $i=n_{k|k-1}+j$, where $j\in\{1,\dots,m_k\}$ and $\iota\in\{1,\dots,2^{j-1}\}$. Then,  under this construction, the $j$-th new Bernoulli component has $h^i_{k|k}=2^{j-1}+1$ local hypotheses, one of which corresponds to non-existence
\begin{equation}\label{eq_non_existence_ber}
  {\cal M}_k^{i,1} = \emptyset, \quad w_{k|k}^{i,1} = 1, \quad r_{k|k}^{i,1} = 0,
\end{equation}
and the others ($a^i = \iota+1,\iota\in \{1,\dots,2^{j-1}\}$), corresponding to detection with ${\bf w}_k^{i,\iota} \in \mathcal{S}_i$, are given by
\begin{subequations}\label{eq_newBernoulli2}
  \begin{align}
    {\cal M}_k^{i,a^i} &= \left\{ (k,p):z_k^p\in {\bf w}_k^{i,\iota} \right\},\\
    \ell_{k|k}^{i,\iota} &= \left\langle \lambda_{k|k-1},\ell_k\left({\bf w}_k^{i,\iota}|\cdot\right) \right\rangle,\label{eq_detection_likelihood_undetected}\\
    w_{k|k}^{i,a^i} &= \delta_1\left[\left|{\bf w}_k^{i,\iota}\right|\right]\lambda_k^C\left(z_k^j\right)+ \ell_{k|k}^{i,\iota},\\
    r_{k|k}^{i,a^i} &= \frac{\ell_{k|k}^{i,\iota}}{w_{k|k}^{i,a^i}},\\
    f_{k|k}^{i,a^i}(x) &= \frac{\ell_k\left({\bf w}_k^{i,\iota}|x\right)\lambda_{k|k-1}(x)}{\ell_{k|k}^{i,\iota}}.
  \end{align}
\end{subequations}
\label{pmbm_update}
\end{proposition}

In \eqref{eq_newBernoulli2}, under the local hypothesis with $\mathbf{w}_k^{i,\iota} = \{z_k^j\}$, the Bernoulli component accounts for the possibility that the measurement $z_k^j$ is either clutter or generated by a newly detected object. If $|{\bf w}_k^{i,\iota}|>1$, the local hypothesis corresponds to an object with existence probability one.

\begin{remark}
The local hypothesis structure in \eqref{eq_new_local_meas} depends on the ordering of measurements. Thus, the ordered construction partitions all non-empty measurement subsets according to their largest indices. As shown in \cite[Theorem 1]{xia2021poisson}, this construction is equivalent, at the global hypothesis level, to the alternative construction in which each non-empty measurement subset creates a new Bernoulli component. Consequently, changing the measurement ordering reorganizes the local hypotheses for newly detected objects, but does not change the represented global hypotheses or the resulting PMBM density.

For the PMBM implementations, measurement ordering therefore affects only the bookkeeping and canonical indexing of local hypotheses for newly detected objects. If a track-oriented PMB approximation is subsequently performed, the ordering may affect which local hypotheses are grouped and merged because the new Bernoulli components are indexed differently \cite{xia2021poisson}. Though PMB approximation is an interesting extension, it is not considered in the proposed implementations.
\label{rem:measurement_ordering}
\end{remark}

\section{Joint Posterior with Auxiliary Variables}

This section constructs the joint posterior used by the blocked Gibbs samplers in three steps. First, we formulate the PMBM density on an augmented object state space by assigning an auxiliary track index to each potential object. Second, we introduce the global hypothesis as an explicit random variable and derive its joint posterior with the set of objects. Finally, we introduce the current measurement-oriented association vector and factorize the joint posterior into terms associated with individual potential objects and measurements. In the resulting representation, a previous global hypothesis and a current measurement association vector jointly determine an updated global hypothesis.

\subsection{PMBM Density with Auxiliary Variables}

We augment the single object space by introducing an auxiliary variable $u\in\mathbb{U}_{k|k^\prime}=\{0,1,\dots,n_{k|k^\prime}\}$, so that each augmented object state $(u,x) \in \mathbb{U}_{k|k^\prime}\times\mathcal{X}$. The auxiliary variable $u$ can be interpreted as a track index: for $u=i\in\{1,\dots,n_{k|k^\prime}\}$, $x$ represents the state of the $i$-th potential object (also referred to as track in \cite{pmbmpoint}), and for $u=0$, $x$ represents an undetected object. A set of objects equipped with track indices is then denoted as $\widetilde{\mathbf{x}}_k\in\mathcal{F}(\mathbb{U}_{k|k^\prime}\times\mathcal{X})$.  Given a PMBM density $f_{k|k^\prime}(\cdot)$ of the form \eqref{eq_pmbm}, the PMBM density on the augmented space $\mathcal{F}(\mathbb{U}_{k|k^\prime}\times\mathcal{X})$ is given by \cite{garcia2020trajectory}
\begin{align}
    {f}_{k|k^\prime}(\widetilde{\mathbf{x}}_k) &=  {f}_{k|k^\prime}^p\left(\widetilde{\mathbf{x}}_k^u\right) {f}_{k|k^\prime}^{mbm}\left(\widetilde{\mathbf{x}}^d_k\right),\label{eq_pmbm_augmented}\\
    {f}_{k|k^\prime}^p\left(\widetilde{\mathbf{x}}_k^u\right) &= e^{-\left\langle {\lambda}_{k|k^\prime},1\right\rangle}\left[ {\lambda}_{k|k^\prime}(\cdot) \right]^{\widetilde{\mathbf{x}}_k^u},\label{eq_ppp_augmented}\\
    {f}_{k|k^\prime}^{mbm}\left(\widetilde{\mathbf{x}}^d_k\right) &= \sum_{a\in\mathcal{A}_{k|k^\prime}}w^a_{k|k^\prime}\prod_{i=1}^{n_{k|k^\prime}}{f}^{i,a^i}_{k|k^\prime}\left(\widetilde{\mathbf{x}}_k^i\right),\label{eq_mbm_augmented}
\end{align}
where $\widetilde{\mathbf{x}}_k=\widetilde{\mathbf{x}}_k^u\uplus\widetilde{\mathbf{x}}_k^d$ and $\widetilde{\mathbf{x}}_k^d=\uplus_{i=1}^{n_{k|k^\prime}}\widetilde{\mathbf{x}}_k^i$, with $\widetilde{\mathbf{x}}_k^u=\{(u,x)\in\widetilde{\mathbf{x}}_k:u=0\}$ and $\widetilde{\mathbf{x}}_k^i=\{(u,x)\in\widetilde{\mathbf{x}}_k:u=i\}$. In \eqref{eq_ppp_augmented}, the Poisson intensity is
\begin{equation}
    {\lambda}_{k|k^\prime}(u,x) = \delta_0[u]\lambda_{k|k^\prime}(x),
\end{equation}
and in \eqref{eq_mbm_augmented}, the $i$-th augmented Bernoulli density under local hypothesis $a^i\in\{1,\dots,h^i_{k|k^\prime}\}$ is
\begin{equation}
    {f}^{i,a^i}_{k|k^\prime}\left(\widetilde{\mathbf{x}}_k^i\right) = \begin{cases}
        1 - r_{k|k^\prime}^{i,a^i} & \widetilde{\mathbf{x}}_k^i = \emptyset \\
        \delta_i[u]r_{k|k^\prime}^{i,a^i} f^{i,a^i}_{k|k^\prime}(x) & \widetilde{\mathbf{x}}_k^i = \{(u,x)\} \\
        0 & \text{otherwise}.
    \end{cases}\label{eq_bernoulli_augmented}
\end{equation}
In contrast to the original PMBM density \eqref{eq_pmbm}, the augmented PMBM density in \eqref{eq_pmbm_augmented} does not require summation over set partitions. This is because the auxiliary variables uniquely decompose $\widetilde{\mathbf{x}}_k$ into $\widetilde{\mathbf{x}}_k^u$, $\widetilde{\mathbf{x}}_k^1,\dots,\widetilde{\mathbf{x}}_k^{n_{k|k^\prime}}$, and only this decomposition yields a non-zero density~\cite{garcia2020trajectory}. If we integrate out the auxiliary variable $u$ for each object, we recover the original PMBM density \eqref{eq_pmbm} \cite{garcia2020trajectory}. We also note that in $\widetilde{\mathbf{x}}_k$ there can be multiple objects with $u=0$, but at most one object with $u=i\in\{1,\dots,n_{k|k^\prime}\}$, for a non-zero density of \eqref{eq_pmbm_augmented}.

\subsection{Joint Posterior of Sets of Objects and Global Hypotheses}

We define the joint posterior over the augmented set $\widetilde{\mathbf{x}}_k$ of objects and the global hypothesis $a$ on the space $\mathcal{F}(\mathbb{U}_{k|k}\times\mathcal{X})\times\mathcal{A}_{k|k}$ as
\begin{align}
    {f}_{k|k}\left(\widetilde{\mathbf{x}}_k,a\right) &= {f}_{k|k}^p\left(\widetilde{\mathbf{x}}_k^u\right) w^a_{k|k} \prod_{i=1}^{n_{k|k}}{f}^{i,a^i}_{k|k}\left(\widetilde{\mathbf{x}}_k^i\right),\nonumber \\
    &\propto {f}_{k|k}^p\left(\widetilde{\mathbf{x}}_k^u\right) \prod_{i=1}^{n_{k|k}} w^{i,a^i}_{k|k} {f}^{i,a^i}_{k|k}\left(\widetilde{\mathbf{x}}_k^i\right),\nonumber \\
    &\triangleq {f}_{k|k}^p\left(\widetilde{\mathbf{x}}_k^u\right) \prod_{i=1}^{n_{k|k}} {g}^{i,a^i}_{k|k}\left(\widetilde{\mathbf{x}}_k^i\right),
    \label{eq_joint_posterior}
\end{align}
where, by Proposition~\ref{pmbm_update}, the terms ${g}^{i,a^i}_{k|k}(\cdot)$ can be obtained for each Bernoulli component $i\in\{1,\dots,n_{k|k}\}$ and each local hypothesis $a^i\in\{1,\dots,h^i_{k|k}\}$ as follows.

Specifically, we have that, for misdetection hypotheses, with $i\in \{1,\dots,n_{k|k-1}\}$ and $a^i \in \{1,\dots,h^i_{k | k-1}\}$,
\begin{align}
  &{g}^{i,a^i}_{k|k}\left(\widetilde{\mathbf{x}}^i_k\right)= \nonumber \\
  & \begin{cases}
    \delta_i[u]w^{i,a^i}_{k|k-1}r^{i,a^i}_{k|k-1}\ell_k(\emptyset|x)f^{i,a^i}_{k|k-1}(x) & \widetilde{\mathbf{x}}^i_k = \{(u,x)\} \\
    w^{i,a^i}_{k|k-1}\left(1-r^{i,a^i}_{k|k-1}\right) & \widetilde{\mathbf{x}}^i_k = \emptyset \\
    0 & \text{otherwise},
  \end{cases}
\end{align}
and for detection hypotheses, where $a^i=\tilde{a}^i+h^i_{k|k-1}j$ with $\tilde{a}^i\in\{1,\dots,h^i_{k|k-1}\}$ and $j\in\{1,\dots,2^{m_k}-1\}$,
\begin{align}
  &{g}^{i,a^i}_{k|k}\left(\widetilde{\mathbf{x}}^i_k\right)= \nonumber\\
  &\begin{cases}
     \delta_i[u]w^{i,\tilde{a}^i}_{k|k-1}r^{i,\tilde{a}^i}_{k|k-1}\ell_k\left({\bf w}_k^j|x\right)f_{k|k-1}^{i,\tilde{a}^i}(x) & \widetilde{\mathbf{x}}_k^i = \{(u,x)\} \\
     0 & \text{otherwise}.
  \end{cases}
\end{align}
For the non-existence hypothesis for new Bernoulli components, with $i = n_{k|k-1}+j,j\in\{1,\dots,m_k\}$,
\begin{equation}
  {g}^{i,1}_{k|k}\left(\widetilde{\mathbf{x}}^i_k\right) = \begin{cases}
    1 & \widetilde{\mathbf{x}}^i_k = \emptyset \\ 
    0 & \text{otherwise},
  \end{cases}
\end{equation}
and for the existence hypotheses for new Bernoulli components, with $a^i = \iota+1$, $\iota\in \{1,\dots,2^{j-1}\}$
\begin{equation}
  {g}^{i,a^i}_{k|k}\left(\widetilde{\mathbf{x}}^i_k\right) 
  = \begin{cases}
    \delta_i[u]\ell_k\left({\bf w}_k^{i,\iota}|x\right)\lambda_{k|k-1}(x) & \widetilde{\mathbf{x}}^i_k = \{(u,x)\} \\ 
   \delta_1\left[\left|{\bf w}_k^{i,\iota}\right|\right]\lambda_k^C\left(z_k^j\right) & \widetilde{\mathbf{x}}^i_k = \emptyset \\
    0 & \text{otherwise}.
  \end{cases}
\end{equation}

\subsection{Joint Posterior with Data Association Variables}

Using the factorized form of the Poisson set density for object-generated measurements in \eqref{eq_ppp_meas}, the joint posterior in \eqref{eq_joint_posterior} can be reformulated as a product of factors associated with individual measurement assignments. To represent this posterior in terms of explicit data association variables, we introduce the measurement-oriented association vector
\begin{equation}
    \beta_k=[\beta_k^1,\dots,\beta_k^{m_k}]^T,
\end{equation}
where $\beta_k^j$ indicates the Bernoulli index to which measurement $z_k^j$ is assigned. According to the PMBM hypothesis structure in Proposition~\ref{pmbm_update}, its support is
\begin{equation}
    \beta_k^j\in\{1,\dots,n_{k|k-1},n_{k|k-1}+j,\dots,n_{k|k}\}.
\end{equation}
Therefore, measurement $z_k^j$ can be assigned to any existing Bernoulli component or to a new Bernoulli component whose index is no smaller than $n_{k|k-1}+j$.

Let $a\in\mathcal{A}_{k|k-1}$ denote a previous global hypothesis and let $\widetilde{a}\in\mathcal{A}_{k|k}$ denote an updated global hypothesis. Given $\widetilde{a}$, its measurement association vector is obtained by setting $\beta_k^j=i$ whenever $(k,j)\in\mathcal{M}_k^{i,\widetilde{a}^i}$. Conversely, the pair $(a,\beta_k)$ uniquely determines $\widetilde{a}$: the previous global hypothesis specifies the local hypotheses before processing the current measurements, while $\beta_k$ specifies how the current measurements update these local hypotheses. The reverse correspondence also holds because $\widetilde{a}$ retains the previous association history and the current measurement assignments. We then have the one-to-one correspondence $(a,\beta_k)\longleftrightarrow\widetilde{a}$. Hereafter, $a$ and $\widetilde{a}$ consistently denote the previous and updated global hypotheses, respectively.

\begin{example}
  Consider a previous global hypothesis with two Bernoulli components and current measurement set $\mathbf{z}_k=\{z_k^1,z_k^2\}$. In this case, two new Bernoulli components are created with indices $i=3$ and $i=4$. Suppose that the association vector is $\beta_k=[1,4]^T$, which assigns $z_k^1$ to Bernoulli component $1$ and $z_k^2$ to Bernoulli component $4$. Consequently, in the updated global hypothesis $\widetilde{a}$, Bernoulli component $1$ is updated using single measurement $\{z_k^1\}$, Bernoulli component $2$ has a misdetection local hypothesis, Bernoulli component $3$ has its non-existence local hypothesis, and Bernoulli component $4$ has the local hypothesis generated by $\{z_k^2\}$. As another example, association $\beta_k=[4,4]^T$ assigns both measurements $\{z_k^1,z_k^2\}$ to Bernoulli component $4$, thereby representing the hypothesis that they were generated by the same newly detected object. 
\end{example}

Using the alternative representation of global hypotheses with the factorized form of \eqref{eq_ppp_meas}, the joint posterior in \eqref{eq_joint_posterior} admits a further decomposition given by
\begin{align}
  &{f}_{k|k}\left(\widetilde{\mathbf{x}}_k,a,\beta_k\right)\nonumber \\ 
  &= {f}_{k|k}^p\left(\widetilde{\mathbf{x}}_k^u\right) {f}_{k|k}^d\left(\widetilde{\mathbf{x}}_k^d,a,\beta_k\right) \label{eq_joint_mbm_factorization}\\
  &\propto {f}_{k|k}^p\left(\widetilde{\mathbf{x}}_k^u\right) \prod_{i=1}^{n_{k|k-1}}w^{i,a^i}_{k|k-1} \left[\underline{f}^{i,a^i}_{k|k-1}\left(\widetilde{\mathbf{x}}_k^i\right)\prod_{j=1}^{m_k}\underline{s}^i_k\left(\widetilde{\mathbf{x}}_k^i,\beta_k^j; z_k^j\right)\right] \nonumber \\
  & ~~~\times \prod_{i = n_{k|k-1}+1}^{n_{k|k}}\left[\overline{f}^i_{k|k-1}\left(\widetilde{\mathbf{x}}_k^i\right)\overline{s}^i_k\left(\widetilde{\mathbf{x}}_k^{i},\beta_k^{i-n_{k|k-1}};z_k^{i-n_{k|k-1}}\right)\right. \nonumber\\
  &~~~~~~~~~~~~~~~~~~~~\times \left. \prod_{j=1}^{i-n_{k|k-1}-1}\underline{s}^i_k\left(\widetilde{\mathbf{x}}_k^i,\beta_k^j;z_k^j\right)\right] \label{eq_joint_pmbm_factorization1}\\
  &= {f}_{k|k}^p\left(\widetilde{\mathbf{x}}_k^u\right)w^{a}_{k|k-1}\prod_{i=1}^{n_{k|k-1}}\underline{f}^{i,a^i}_{k|k-1}\left(\widetilde{\mathbf{x}}_k^i\right)\nonumber\\
  &\times\prod_{i = n_{k|k-1}+1}^{n_{k|k}}\overline{f}^i_{k|k-1}\left(\widetilde{\mathbf{x}}_k^i\right) \nonumber \\
  &\times  \prod_{j=1}^{m_k}\left[\overline{s}^{n_{k|k-1}+j}_k\left(\widetilde{\mathbf{x}}_k^{n_{k|k-1}+j},\beta_k^j;z_k^j\right)\prod_{i \in \mathcal{I}_k^j}\underline{s}^i_k\left(\widetilde{\mathbf{x}}_k^i,\beta_k^j;z_k^j\right)\right] \label{eq_joint_pmbm_factorization2}
\end{align}
where $a \in \mathcal{A}_{k|k-1}$, $\mathcal{I}_k^j \triangleq i \in \{1,\dots,n_{k|k-1}\} \cup \{n_{k|k-1}+j+1,\dots,n_{k|k}\}$, and the new factors are
\begin{align}
  &\underline{f}^{i,a^i}_{k|k-1}\left(\widetilde{\mathbf{x}}_k^i\right) \nonumber \\
  &= \begin{cases}
    \delta_i[u]r^{i,a^i}_{k|k-1}e^{-\gamma_k(x)}f^{i,a^i}_{k|k-1}(x) & \widetilde{\mathbf{x}}_k^i = \{(u,x)\} \\
    1-r^{i,a^i}_{k|k-1} & \widetilde{\mathbf{x}}_k^i  = \emptyset \\
    0 & \text{otherwise},
  \end{cases} \label{eq_existing_prior}\\
  &\overline{f}^i_{k|k-1}\left(\widetilde{\mathbf{x}}_k^i\right) = \begin{cases}
    \delta_i[u]e^{-\gamma_k(x)}\lambda_{k|k-1}(x) & \widetilde{\mathbf{x}}_k^i = \{(u,x)\} \\
    1 & \widetilde{\mathbf{x}}_k^i  = \emptyset \\
    0 & \text{otherwise},
  \end{cases} \label{eq_new_prior}\\
  &\underline{s}^i_k\left(\widetilde{\mathbf{x}}_k^i,\beta_k^j;z_k^j\right) \nonumber \\
  &= \begin{cases}
    \delta_i[u]\gamma_k(x)\ell_k\left(z_k^j | x\right) & \widetilde{\mathbf{x}}_k^i = \{(u,x)\}, \beta_k^j = i \\
    1 & \beta_k^j \neq i \\
    0 & \text{otherwise},
  \end{cases} \label{eq_s_underline}\\
  &\overline{s}^i_k\left(\widetilde{\mathbf{x}}_k^i,\beta_k^j;z_k^j\right) \nonumber \\
  &= \begin{cases}
    \delta_i[u]\gamma_k(x)\ell_k\left(z_k^j | x\right) & \widetilde{\mathbf{x}}_k^i = \{(u,x)\}, \beta_k^j = i \\
    \lambda^C_k\left(z_k^j\right) & \widetilde{\mathbf{x}}_k^i = \emptyset, \beta_k^j = i \\
    1 & \widetilde{\mathbf{x}}_k^i = \emptyset, \beta_k^j \neq i \\
    0 & \text{otherwise}.
  \end{cases} \label{eq_s_overline}
\end{align}

In \eqref{eq_joint_pmbm_factorization1}, the product over $i\in\{1,\dots,n_{k|k-1}\}$ corresponds to the contribution of previously existing objects, whereas the product over $i\in\{n_{k|k-1}+1,\dots,n_{k|k}\}$ is for newly detected objects. Together, these two products characterize the joint posterior of the set $\widetilde{\mathbf{x}}_k^d=\uplus_{i=1}^{n_{k|k}}\widetilde{\mathbf{x}}_k^i$ of detected objects. Since a potential object may generate multiple measurements, the factorization of the PPP measurement likelihood in \eqref{eq_ppp_meas} leads directly to the factorized structure in \eqref{eq_joint_pmbm_factorization1}.

The different factors in \eqref{eq_joint_pmbm_factorization1} are rearranged in \eqref{eq_joint_pmbm_factorization2}. In this expression, the first two products over $\underline{f}^{i,a^i}_{k|k-1}(\cdot)$ and $\overline{f}^i_{k|k-1}(\cdot)$ collect the predicted and prior terms for existing and newly detected objects, respectively, whereas the final product over $\overline{s}^i_k(\cdot,\cdot)$ and $\underline{s}^i_k(\cdot,\cdot)$ contains the measurement-update factors. The factor $\overline{s}^i_k(\cdot,\cdot)$ is specific to newly detected objects, while $\underline{s}^i_k(\cdot,\cdot)$ applies to both existing and newly detected objects. Moreover, the second line of \eqref{eq_s_overline} represents the case that measurement $z_k^j$ is clutter, so the associated potential object is absent.

The proposed factorization \eqref{eq_joint_pmbm_factorization2} is related to the factorization in \cite{xia2023trajectory}, but differs in three main aspects. First, \cite{xia2023trajectory} considers PMB filtering, where the predicted density contains a single global hypothesis, whereas the PMBM density considered here contains multiple predicted global hypotheses. The previous global hypothesis $a$ must therefore be explicitly included in the proposed factorization. Second, the factorization in \cite{xia2023trajectory} is for sets of trajectories, whereas the present formulation operates on the set of objects at the current time step. Third, the measurement-oriented association variable treats clutter differently in the two formulations, as explained in Remark~\ref{rem:association_variable}.

\begin{remark}
In the joint posterior with measurement association variables presented in \cite{xia2023trajectory,xia2023efficient}, $\beta_k^j$ can also be 0, which explicitly represents the case that $z_k^j$ is a clutter detection. In contrast, in this paper, $\beta_k^j$ cannot be zero. The hypothesis that $z_k^j$ is a clutter detection is implicitly captured in \eqref{eq_joint_pmbm_factorization2} by the new Bernoulli components via the second line of \eqref{eq_s_overline}. This aligns well with the new Bernoulli densities \eqref{eq_newBernoulli2} and results in a slightly more compact factorization.
\label{rem:association_variable}
\end{remark}

\begin{remark}
The joint posterior in \eqref{eq_joint_pmbm_factorization2} for Poisson birth can be adapted to an MB birth model with modest modifications. Specifically, under the MB birth model, the Poisson intensity of undetected objects becomes zero, and we introduce new Bernoulli birth components in the prediction step. As a result, the factors in \eqref{eq_new_prior} and \eqref{eq_s_overline}, which account for measurement-initiated new Bernoulli components, vanish. Moreover, the Poisson clutter intensity $\lambda_k^C(\cdot)$ in the second line of \eqref{eq_s_overline} can be absorbed into the factor in \eqref{eq_s_underline} by dividing its first line by $\lambda_k^C(\cdot)$ and extending the support of $\beta_k^j$ to include $0$, corresponding to the case where $z_k^j$ is clutter. This also means that the PMBM implementations presented in the following can be easily adapted to MBM.
\end{remark}

\section{PMBM Update Using Blocked Gibbs Sampling}

This section presents an efficient implementation of the PMBM update step using blocked Gibbs sampling, which propagates global hypotheses with high weights by iteratively sampling from the conditional distributions of the previous global hypothesis, the current measurement association variables, and the set of detected objects. In addition, we show that the Bernoulli existence variables can be marginalized out in sampling the set of detected objects, yielding a collapsed sampler with improved efficiency. Finally, we describe how the computational efficiency of blocked Gibbs sampling can be further enhanced through a track-oriented implementation.

\subsection{Blocked Gibbs Sampling}
\label{sec_full_Gibbs}
Maintaining tractable computational complexity in the PMBM update step relies on truncating global hypotheses in the MBM with negligible weights (also called global hypothesis pruning). In this paper, we aim to approximate the MBM using samples of the global hypotheses. Instead of directly sampling data associations as in \cite{fatemi2017poisson,soextended}, we sample the joint posterior distribution over set of detected objects $\widetilde{\mathbf{x}}_k^d$ and global hypothesis $(a,\beta_k)$, with dual object-oriented and measurement-oriented representations, using Gibbs sampling. In addition, since the posterior density of undetected objects ${f}_{k|k}^p\left(\widetilde{\mathbf{x}}_k^u\right)$ does not depend on any global hypotheses $(a,\beta_k)$, it is sufficient to target the joint posterior ${f}_{k|k}^d\left(\widetilde{\mathbf{x}}_k^d,a,\beta_k\right)$ \eqref{eq_joint_mbm_factorization} for detected objects.

\begin{figure}[!t]
    \centering
    \includegraphics[width=\columnwidth]{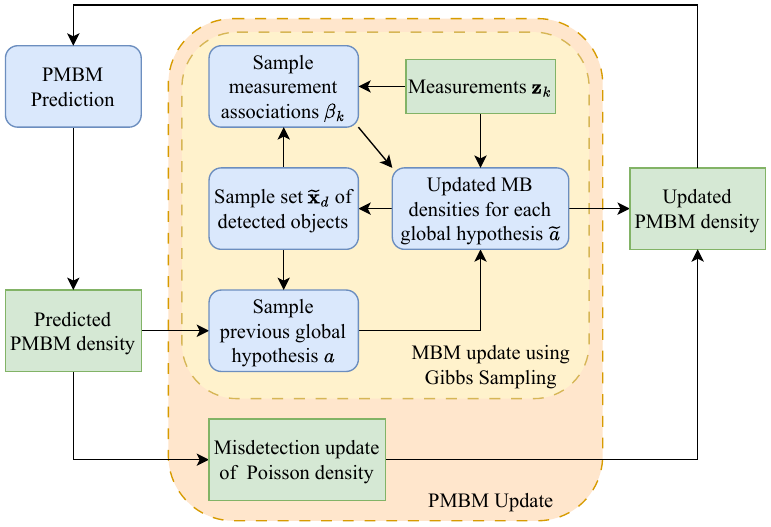}
    \caption{Diagram of the extended object PMBM filter based on blocked Gibbs sampling.}
    \label{fig_pmbm_diagram}
\end{figure}

Specifically, to draw samples from the joint posterior ${f}_{k|k}^d\left(\widetilde{\mathbf{x}}_k^d,a,\beta_k\right)$, we develop a blocked Gibbs sampler that iteratively samples the following conditional distributions: 
\begin{enumerate}
  \item Conditional distribution of previous global hypothesis ${f}_{k|k}\left(a | \widetilde{\mathbf{x}}^d_k,\beta_k\right) = {f}_{k|k}\left(a | \widetilde{\mathbf{x}}^d_k\right)$.
  \item Conditional distribution of data association variable ${f}_{k|k}\left(\beta_k | \widetilde{\mathbf{x}}^d_k,a\right) = {f}_{k|k}\left(\beta_k | \widetilde{\mathbf{x}}^d_k\right)$.
  \item Conditional distribution of set $\widetilde{\mathbf{x}}^d_k$ of detected objects ${f}_{k|k}\left(\widetilde{\mathbf{x}}^d_k | a,\beta_k\right)$.
\end{enumerate}
The diagram of the extended object PMBM filter using blocked Gibbs sampling is illustrated in Fig. \ref{fig_pmbm_diagram}. Next, we derive the expressions of these conditional distributions, leveraging the factorization of ${f}_{k|k}^d(\widetilde{\mathbf{x}}_k^d,a,\beta_k)$ in \eqref{eq_joint_mbm_factorization}.

\subsubsection{Sampling previous global hypotheses}

It can be observed from \eqref{eq_joint_pmbm_factorization2} that the previous global hypothesis $a\in \mathcal{A}_{k|k-1}$ only appears in its weight $w^a_{k|k-1}$ and the first product over $\underline{f}^{i,a^i}_{k|k-1}(\cdot)$ \eqref{eq_existing_prior}. Therefore, the conditional distribution of $a$ can be expressed as
\begin{align}
  {f}_{k|k}\left(a | \widetilde{\mathbf{x}}^d_k,\beta_k\right) &= \frac{{f}_{k|k}^d\left(\widetilde{\mathbf{x}}_k^d,a,\beta_k\right)}{\sum_{a\in\mathcal{A}_{k|k-1}}{f}_{k|k}^d\left(\widetilde{\mathbf{x}}_k^d,a,\beta_k\right)} \nonumber\\
  &\propto w^a_{k|k-1}\prod_{i=1}^{n_{k|k-1}}\underline{f}^{i,a^i}_{k|k-1}\left(\widetilde{\mathbf{x}}_k^i\right).\label{eq_sample_a}
\end{align}
This means that the conditional distribution of $a$ only depends on the set of existing objects $\uplus_{i=1}^{n_{k|k-1}}\widetilde{\mathbf{x}}_k^i$. 

\subsubsection{Sampling data associations}

The conditional distribution of data association variable $\beta_k$ also has a simple form since it only appears in the last big product of \eqref{eq_joint_pmbm_factorization2}. Thus, we have
\begin{align}
  &{f}_{k|k}\left(\beta_k | \widetilde{\mathbf{x}}^d_k,a\right) = \frac{{f}_{k|k}^d\left(\widetilde{\mathbf{x}}_k^d,a,\beta_k\right)}{\sum_{\beta_k}{f}_{k|k}^d\left(\widetilde{\mathbf{x}}_k^d,a,\beta_k\right)} \nonumber\\
  &\propto \prod_{j=1}^{m_k}\left[\overline{s}_k^{n_{k|k-1}+j}\left(\widetilde{\mathbf{x}}_k^{n_{k|k-1}+j},\beta_k^j;z_k^j\right)  \prod_{i \in \mathcal{I}_k^j}\underline{s}^i_k\left(\widetilde{\mathbf{x}}_k^i,\beta_k^j;z_k^j\right)\right]\label{eq_sample_beta}
\end{align}
which is proportional to a product of factors, and each factor only depends on a single measurement association variable $\beta_k^j$. Therefore, we can sample each $\beta_k^j$ independently according to
\begin{align}
  &{f}_{k|k}\left(\beta_k^j | \widetilde{\mathbf{x}}^d_k,a\right) \nonumber \\
  &\propto \overline{s}_k^{n_{k|k-1}+j}\left(\widetilde{\mathbf{x}}_k^{n_{k|k-1}+j},\beta_k^j;z_k^j\right)\prod_{i \in \mathcal{I}_k^j}\underline{s}^i_k\left(\widetilde{\mathbf{x}}_k^i,\beta_k^j;z_k^j\right) \label{eq_beta_k_j}.
\end{align}
We note that the right-hand side of \eqref{eq_beta_k_j} does not depend on $a$. In addition, plugging \eqref{eq_s_underline} and \eqref{eq_s_overline} into \eqref{eq_beta_k_j}, the probability of $\beta^j_k=i$ only depends on $\widetilde{\mathbf{x}}_k^i$ and $\widetilde{\mathbf{x}}^{n_{k|k-1}+j}_k$ and thus can be further simplified, for $i\in\{1,\dots,n_{k|k-1}\}$,
\begin{align}
  &{f}_{k|k}\left(\beta_k^j = i | \widetilde{\mathbf{x}}^i_k,\widetilde{\mathbf{x}}^{n_{k|k-1}+j}_k\right) \nonumber \\
  &\propto \begin{cases}
    \delta_i[u]\gamma_k(x)\ell_k\left(z_k^j | x\right) & \widetilde{\mathbf{x}}_k^i = \{(u,x)\}, {\mathbf{x}}^{n_{k|k-1}+j}_k = \emptyset \\
    0 & \text{otherwise},
  \end{cases}\label{eq_sample_beta1}
\end{align}
and for $i\in\{n_{k|k-1}+1,\dots,n_{k|k}\}$,
\begin{align}
  &{f}_{k|k}\left(\beta_k^j = i | \widetilde{\mathbf{x}}^i_k,\widetilde{\mathbf{x}}^{n_{k|k-1}+j}_k\right) \nonumber \\
  &\propto \begin{cases}
    \delta_i[u]\gamma_k(x)\ell_k\left(z_k^j | x\right) & \substack{\widetilde{\mathbf{x}}_k^i = \{(u,x)\}, {\mathbf{x}}^{n_{k|k-1}+j}_k = \emptyset \\ j < i - n_{k|k-1}} \\
    \delta_i[u]\gamma_k(x)\ell_k\left(z_k^j | x\right) & \widetilde{\mathbf{x}}_k^i = \{(u,x)\}, j = i - n_{k|k-1} \\
    \lambda_k^C\left(z_k^j\right) & \widetilde{\mathbf{x}}_k^i = \emptyset, j = i - n_{k|k-1} \\
    0 & \text{otherwise},
  \end{cases} \label{eq_sample_beta2}
\end{align}
and, we compute the normalization constant by summing over all admissible assignments of the $j$-th measurement, i.e., $\beta_k^j=i$ for $i\in\{1,\dots,n_{k|k}\}$. Therefore, the elements of the association vector $\beta_k$ can be drawn independently from categorical distributions, allowing for efficient parallel sampling.

It is worth emphasizing that the probability of assigning measurement $z_k^j$ to Bernoulli component $i$, i.e., $\beta_k^j=i$, depends jointly on the object sets $\widetilde{\mathbf{x}}_k^i$ and $\widetilde{\mathbf{x}}_k^{n_{k|k-1}+j}$, which corresponds to the new Bernoulli component initiated by measurement $z_k^j$. When $i\neq n_{k|k-1}+j$, the assignment $\beta_k^j=i$ is only possible if $z_k^j$ is not explained by a newly detected object, namely when $\widetilde{\mathbf{x}}_k^{n_{k|k-1}+j}=\emptyset$. If $i=n_{k|k-1}+j$ and $\widetilde{\mathbf{x}}_k^{n_{k|k-1}+j}=\emptyset$, then $z_k^j$ is interpreted as clutter.

\subsubsection{Sampling sets of detected objects}

The conditional distribution of the set $\widetilde{\mathbf{x}}^d_k$ of detected objects can be obtained by performing the PMBM update given in Proposition \ref{pmbm_update}. Let $\widetilde{a}\in\mathcal{A}_{k|k}$ be the global hypothesis in the PMBM posterior resulting from updating the previous global hypothesis $a\in\mathcal{A}_{k|k-1}$ using association $\beta_k$. For each $\widetilde{a}$, we obtain the updated MB density as
\begin{equation}
  {f}_{k|k}\left(\widetilde{\mathbf{x}}^d_k | a,\beta_k\right) =  {f}_{k|k}\left(\widetilde{\mathbf{x}}^d_k | \widetilde{a}\right) =\prod_{i=1}^{n_{k|k}}{f}^{i,\widetilde{a}^i}_{k|k}\left(\widetilde{\mathbf{x}}_k^i\right), \label{eq_set_detected_conditional}
\end{equation}
where each ${f}^{i,\widetilde{a}^i}_{k|k}(\cdot)$ is a Bernoulli set density augmented with track index $u = i$ for $i \in \{1,\dots,n_{k|k}\}$, see \eqref{eq_bernoulli_augmented}. To sample a set of detected objects from \eqref{eq_set_detected_conditional}, we first draw a sample from each Bernoulli component ${f}^{i,\widetilde{a}^i}_{k|k}(\cdot)$, which is either empty or a singleton, and then take their union. Importantly, each updated Bernoulli component can be computed and sampled independently.

\subsubsection{Initialization}\label{sec_initialization}

The proposed blocked Gibbs sampler can be initialized with any previous global hypothesis $a\in\mathcal{A}_{k|k-1}$ and its valid measurement association vector $\beta_k$. Nevertheless, an informed initialization can typically make the sampling algorithm easier to find high-weight global hypotheses. Empirically, we have found the following initialization strategy works well. Specifically, We begin by selecting the previous global hypothesis $a$ with the highest weight $\arg\max_{a} w^a_{k|k-1}$, and the measurement association $\beta_k$ is then determined as follows:
\begin{subequations}
\begin{align}
  \beta_k^j &= 
    i \text{ for } \varphi(j) = i \in \{1,\dots,n_{k|k-1}\}, \\
  \varphi(j) &= \underset{{i\in\{0,1,\dots,n_{k|k-1}\}}}{\arg\max} \ell_{\text{init}}^{j,i},\\
  \ell_{\text{init}}^{j,0} &= \lambda_k^C\left(z_k^j\right) + \left\langle \lambda_{k|k-1},\ell_k\left(\left\{z_k^j\right\}|\cdot\right) \right\rangle,\\
  \ell_{\text{init}}^{j,i} &= r_{k|k-1}^{i,a^i} \left\langle f_{k|k-1}^{i,{a}^i},\ell_k\left(\left\{z_k^j\right\}|\cdot\right) \right\rangle.
\end{align}
\end{subequations}
In other words, for each measurement $z_k^j$, we first select the association index that maximizes the likelihood $\ell_{\text{init}}^{j,i}$. If $\varphi(j)=i\in\{1,\dots,n_{k|k-1}\}$, the measurement is associated with existing Bernoulli component $i$. Measurements not being assigned to any existing Bernoulli component are subsequently clustered using DBSCAN \cite{ester1996density}, where each cluster corresponds to a new Bernoulli component, whose index is assigned based on the largest measurement index within that cluster according to \eqref{eq_new_local_meas}. 

We note that the DBSCAN-based procedure described above is not essential for the proposed sampling method. As demonstrated in Section~\ref{sec_simulation}, it substantially improves the full blocked Gibbs sampler, whereas the approximate collapsed sampler introduced in the next subsection is less sensitive to initialization. Moreover, alternative clustering methods can be used, e.g., the distance partitioning \cite{phdextended2}.

\subsection{Collapsed Gibbs Sampling}

Drawing the set $\widetilde{\mathbf{x}}^d_k$ of detected objects from \eqref{eq_set_detected_conditional} can be viewed as a two-step procedure. For each Bernoulli component $i\in\{1,\dots,n_{k|k}\}$, we first sample its cardinality $e_k^i=|\widetilde{\mathbf{x}}_k^i|\in\{0,1\}$, which serves as the existence indicator of the $i$-th potential object: $e_k^i=1$ if the object exists and $e_k^i=0$ otherwise. The object state is sampled only when $e_k^i=1$. However, when a Bernoulli component has a small probability of existence, this procedure often produces an empty set. This situation arises, e.g., for local hypotheses of new Bernoulli components initiated by a single measurement and for local hypotheses associated with missed detections. To improve the efficiency of the blocked Gibbs sampler, we therefore marginalize out the set cardinality and sample object states for all Bernoulli components with non-zero existence probability. This is usually known as (partially) collapsed Gibbs sampling in the literature \cite{van2008partially}, which is a useful technique to improve the efficiency of Gibbs sampler by marginalizing out some variables from the sampling process.

To make the collapsed Gibbs sampling strategy more explicit, for each Bernoulli component $i\in\{1,\dots,n_{k|k}\}$, we denote its state by $(u^i,x_k^i)$ when the corresponding object exists. The Bernoulli set $\widetilde{\mathbf{x}}_k^i$ can then be parameterized by $(u^i,x_k^i,e_k^i)$, where $e_k^i\in\{0,1\}$ is the existence indicator and $x_k^i$ is a dummy variable whenever $e_k^i=0$. Once the existence variables $e_k^i$ are marginalized out of the sampling procedure, the conditional distribution for drawing $\widetilde{\mathbf{x}}_k^d$, namely \eqref{eq_set_detected_conditional}, simplifies to
\begin{equation}
  {f}_{k|k}\left(\left\{\left(u^i,x^i_k\right)\right\}_{i=1}^{n_{k|k}} | \widetilde{a} \right) =\prod_{i=1}^{n_{k|k}}\delta_i\left[u^i\right]{f}^{i,\widetilde{a}^i}_{k|k}\left(x_k^i\right).\label{eq_set_of_all_samples}
\end{equation}
Note that we do not need to sample Bernoulli components with zero existence probability \eqref{eq_non_existence_ber}, and that \eqref{eq_set_of_all_samples} is in essence the MB density \eqref{eq_set_detected_conditional} with all existence probabilities equal to one. 

Since the existence variables $e_k^i$ are removed from the iterative sampling procedure, they must also be marginalized out in the conditional distributions of the previous global hypothesis \eqref{eq_sample_a} and the current measurement associations \eqref{eq_sample_beta}, conditioned on the set of detected objects $\widetilde{\mathbf{x}}_k^d$. We also note that $e_k^i$ depends on the updated global hypothesis $\widetilde{a}$, which in turn is determined by the previous global hypothesis $a$ and the measurement association vector $\beta_k$ from the preceding sampling iteration. As a result, the conditional distributions of $a$ and $\beta_k$ become ${f}_{k|k}(a \mid \{(u^i,x_k^i)\}_{i=1}^{n_{k|k-1}},\beta_k)$ and ${f}_{k|k}(\beta_k \mid \{(u^i,x_k^i)\}_{i=1}^{n_{k|k-1}},a)$.

Marginalizing out $e_k^i$ therefore increases the dependency between successive samples of the remaining variables, because the corresponding conditional distributions inherit the posterior coupling that was previously mediated by the existence variables. Although this can increase sample autocorrelation \cite{van2008partially}, collapsing may still improve overall sampling efficiency by reducing the dimension of the sampling space and removing the variability associated with explicitly sampling $e_k^i$.

After marginalizing out all the existence variables $e_k^{i}$,  we omit the dependencies of ${f}_{k|k}(a | \{(u^i,x_k^i)\}_{i=1}^{n_{k|k-1}},\beta_k)$ and ${f}_{k|k}(\beta_k | \{(u^i,x_k^i)\}_{i=1}^{n_{k|k-1}},a)$ on $\beta_k$ and $a$, respectively, hereafter for notational brevity. The resulting conditional distributions now have the expressions:
\begin{align}
  &{f}_{k|k}\left(a | \left\{\left(u^i,x_k^i\right)\right\}_{i=1}^{n_{k|k-1}}\right)\nonumber\\ 
  &= \sum_{e_k^{1:n_{k|k-1}}}{f}_{k|k}\left(a | \left\{\left(u^i,x_k^i,e_k^i\right)\right\}_{i=1}^{n_{k|k-1}}\right)p\left(e_k^{1:n_{k|k-1}} \right) \nonumber \\
  &\propto w^a_{k|k-1} \sum_{e_k^{1:n_{k|k-1}}}\prod_{i=1}^{n_{k|k-1}}\underline{f}^{i,a^i}_{k|k-1}\left(u^i,x_k^i,e_k^i\right)p\left(e_k^i \right) \nonumber \\
  &\triangleq w^a_{k|k-1} \widetilde{f}^a_{k|k}\left(\left\{\left(u^i,x_k^i\right)\right\}_{i=1}^{n_{k|k-1}}  \right),\label{eq_marginal_global_hypothesis}
\end{align}
where we write $e_k^{1:n_{k|k-1}} = \left(e_k^1,\dots,e_k^{n_{k|k-1}}\right)$, and
\begin{align}
  &{f}_{k|k}\left(\beta_k | \left\{\left(u^i,x_k^i\right)\right\}_{i=1}^{n_{k|k}}\right) \nonumber\\
  &= \sum_{e_k^{1:n_{k|k}}}{f}_{k|k}\left(\beta_k | \left\{\left(u^i,x_k^i,e_k^i\right)\right\}_{i=1}^{n_{k|k}}\right)p\left(e_k^{1:n_{k|k}}  \right) \nonumber \\
  &= \sum_{e_k^{1:n_{k|k}}}\prod_{j=1}^{m_k}{f}_{k|k}\left(\beta^j_k | \left\{\left(u^i,x_k^i,e_k^i\right)\right\}_{i=1}^{n_{k|k}}\right)\prod_{i=1}^{n_{k|k}}p\left(e_k^i \right),\label{eq_marginal_data_association}
\end{align}
with $p(e_k^i = 1) = r^{i,\widetilde{a}^i}_{k|k}$
and $p(e_k^i = 0) = 1 - r^{i,\widetilde{a}^i}_{k|k}$. 


\subsection{Approximate Collapsed Gibbs Sampling}

Marginalizing the Bernoulli existence variables introduces additional dependence among the remaining variables. Consequently, evaluating the exact collapsed conditional distributions in \eqref{eq_marginal_global_hypothesis} and \eqref{eq_marginal_data_association} requires summation over all possible configurations of the Bernoulli existence variables. To obtain tractable sampling distributions, we replace these coupled distributions with the following two approximations.

\begin{itemize}
  \item[A1:] The factor $\widetilde{f}^a_{k|k}(\cdot)$ in the conditional distribution of previous global hypothesis in \eqref{eq_marginal_global_hypothesis} is approximated by a product of factors associated with individual existing Bernoulli components:
  \begin{equation}
    \widetilde{f}^a_{k|k}\left(\left\{\left(u^i,x_k^i\right)\right\}_{i=1}^{n_{k|k-1}}\right) \approx \prod_{i=1}^{n_{k|k-1}}q^{i,a^i}\left(u^i,x_k^i\right).
    \label{eq_approx_predicted_global_hypothesis}
  \end{equation}

  \item[A2:] The conditional distribution of the measurement association vector in \eqref{eq_marginal_data_association} is approximated by a product of measurement-wise conditional distributions:
  \begin{align}
    &{f}_{k|k}\left(\beta_k \mid \left\{\left(u^i,x_k^i\right)\right\}_{i=1}^{n_{k|k}}\right) \nonumber\\
    &\qquad\approx \prod_{j=1}^{m_k}q^j\left(\beta_k^j \mid \left\{\left(u^i,x_k^i\right)\right\}_{i=1}^{n_{k|k}}\right).
    \label{eq_approx_meas_assoc}
  \end{align}
\end{itemize}

Approximation A1 removes the dependence among the potential existing objects when sampling the previous global hypothesis, whereas Approximation A2 removes the dependence among the measurement association variables. Both approximations therefore restore component-wise sampling. The factors are computed by minimizing the Kullback-Leibler (KL) divergence between the corresponding exact collapsed distribution to the product-form approximation. Specifically, the optimal solution is given by the product of the corresponding marginal distributions \cite[p.~277]{koller2009probabilistic}. This type of marginal product approximation has been widely used in MOT, e.g., in the classic JPDA filter \cite{fortmann1983sonar,pmbmpoint}.

The required conditional distributions under approximations A1 and A2 are given in the following propositions. Their derivations are provided in Appendix \ref{app:approx_a1} and \ref{app:approx_a2}.

\begin{proposition}
Under Approximation A1, the product-form approximation that minimizes the KL divergence is obtained from the individual component marginals. For Bernoulli component $i\in\{1,\dots,n_{k|k-1}\}$, the resulting factor is
\begin{align}
  q^{i,a^i}\left(u^i,x_k^i\right) &\propto \sum_{e_k^i}\underline{f}^{i,a^i}_{k|k-1}\left(u^i,x_k^i,e_k^i\right)p\left(e_k^i\right) \nonumber\\
  &= \delta_i\left[u^i\right]r^{i,a^i}_{k|k-1}e^{-\gamma_k\left(x_k^i\right)}f^{i,a^i}_{k|k-1}\left(x_k^i\right)r^{i,\widetilde{a}^i}_{k|k} \nonumber\\
  &\quad+ \left(1-r^{i,a^i}_{k|k-1}\right)\left(1-r^{i,\widetilde{a}^i}_{k|k}\right).
  \label{eq_predicted_global_hypo_marginal_existence}
\end{align}
Consequently, the approximate conditional distribution used to sample the previous global hypothesis is
\begin{align}
  &{f}_{k|k}\left(a \mid \left\{\left(u^i,x_k^i\right)\right\}_{i=1}^{n_{k|k-1}}\right) \propto\nonumber\\
  & w^a_{k|k-1} \prod_{i=1}^{n_{k|k-1}}\left( \delta_i\left[u^i\right]r^{i,a^i}_{k|k-1}e^{-\gamma_k\left(x_k^i\right)} \right.  f^{i,a^i}_{k|k-1}\left(x_k^i\right)r^{i,\widetilde{a}^i}_{k|k} \nonumber\\
  &~~~~~~~~~~~~~~~~~\left.+ \left(1-r^{i,a^i}_{k|k-1}\right)\left(1-r^{i,\widetilde{a}^i}_{k|k}\right)\right).
  \label{eq_collapsed_sample_a}
\end{align}
\label{prop:approx_a1}
\end{proposition}

\begin{proposition}
Under Approximation A2, the product-form approximation minimizing the KL divergence is given by the product of the measurement-wise marginal distributions. For Bernoulli component $i\in\{1,\dots,n_{k|k-1}\}$, the approximate probability of being associated with $z_k^j$ is
\begin{align}
  &q^j\left(\beta_k^j=i \mid \left(u^i,x_k^i\right),\left(u^{i^\prime},x_k^{i^\prime}\right)\right) \nonumber\\
  &\propto \delta_i\left[u^i\right]\gamma_k\left(x_k^i\right)\ell_k\left(z_k^j \mid x_k^i\right)r^{i,\widetilde{a}^i}_{k|k}\left(1-r^{i^\prime,\widetilde{a}^{i^\prime}}_{k|k}\right),
  \label{eq_collapsed_sample_beta1}
\end{align}
where $i^\prime=n_{k|k-1}+j$.
For a new Bernoulli component $i\in\{n_{k|k-1}+1,\dots,n_{k|k}\}$, the corresponding approximate probability is
\begin{align}
  &q^j\left(\beta_k^j=i \mid \left(u^i,x_k^i\right),\left(u^{i^\prime},x_k^{i^\prime}\right)\right) \propto \nonumber\\
  &\begin{cases}
    \delta_i\left[u^i\right]\gamma_k\left(x_k^i\right)\ell_k\left(z_k^j \mid x_k^i\right)r^{i,\widetilde{a}^i}_{k|k}\\
    \quad\times\left(1-r^{i^\prime,\widetilde{a}^{i^\prime}}_{k|k}\right), & j<i-n_{k|k-1},\\
    \delta_i\left[u^i\right]\gamma_k\left(x_k^i\right)\ell_k\left(z_k^j \mid x_k^i\right)r^{i,\widetilde{a}^i}_{k|k}\\
    \quad+\lambda_k^C\left(z_k^j\right)\left(1-r^{i,\widetilde{a}^i}_{k|k}\right), & j=i-n_{k|k-1},\\
    0, & j>i-n_{k|k-1}.
  \end{cases}
  \label{eq_collapsed_sample_beta2}
\end{align}
These probabilities are normalized by summing over the admissible values of $\beta_k^j$.
\label{prop:approx_a2}
\end{proposition}

Propositions~\ref{prop:approx_a1} and \ref{prop:approx_a2} show that, after applying Approximations A1 and A2, the previous global hypothesis and the individual measurement associations can still be sampled efficiently. In particular, the individual variables $\beta_k^j$ can be sampled independently and therefore in parallel. These conditional distributions are approximate: unless A1 and A2 hold exactly, the resulting collapsed sampler does not target the exact joint posterior in \eqref{eq_joint_mbm_factorization}. Their role is instead to provide an efficient mechanism for exploring the global hypothesis space.

\subsubsection{Practical aspects}

A practical subtlety may arise when sampling the local hypotheses of new Bernoulli components. According to Proposition~\ref{pmbm_update}, the local hypothesis structure of the $j$-th new Bernoulli component enforces that it is associated with either an empty set or a set of measurements whose maximum index is $j$, see \eqref{eq_new_local_meas}. Under the parallel sampling of measurement associations, this canonical structure can be temporarily violated. For example, the $j$-th measurement may be reassigned to another Bernoulli component while the $j$-th new Bernoulli component receives a measurement with index $<j$. Although such data association still corresponds to an equivalent unordered collection of new Bernoulli components, it becomes inconsistent with their canonical local hypothesis structure. For such cases, we thus apply a remapping step that modifies the sampled measurement associations to recover the canonical local hypothesis structure, ensuring that equivalent unordered MB densities are not treated as distinct global hypotheses.

Approximations A1 and A2 replace the exact conditional distributions of $a$ and $\beta_k$ with KL-optimal product-form approximations. Consequently, the collapsed Gibbs sampler does not generally target the exact joint posterior in \eqref{eq_joint_mbm_factorization}. These approximations are used only to explore the global hypothesis space; for each sampled pair $(a,\beta_k)$, the updated local hypotheses and weights are computed using the standard PMBM update in Proposition~\ref{pmbm_update}. Therefore, every retained updated global hypothesis is valid, and the resulting density remains a PMBM whose MBM contains a truncated set of global hypotheses. PMBM conjugacy concerns preservation of the PMBM density form under prediction and update, rather than tractable enumeration of all the global hypotheses. The truncation error can be characterized by the discarded global hypothesis weights: the $L_1$ error of the unnormalized truncated density equals the total discarded weight, while the $L_1$ error after normalization is bounded by twice this quantity \cite[Section V-D]{pmbmextended2}.

Finally, we notice that a full sweep of the collapsed Gibbs sampler, which amounts to sampling of the three blocks of variables in each iteration, may be more computationally expensive per iteration than the full Gibbs sampler, since it typically requires drawing more object states and also considering a larger set of possible measurement associations. Nevertheless, collapsing can typically focus the sampling on high-likelihood association patterns and hence requires substantially fewer iterations, resulting in lower overall runtime and improved estimation accuracy, as demonstrated in the simulations.

\begin{algorithm}[t]
  \small
\caption{PMBM update with blocked Gibbs sampling}
\label{alg_full_gibbs}
\begin{algorithmic}[1]
\Require
Predicted PMBM density $f_{k|k-1}(\widetilde{\mathbf{x}}_k)$, number of Gibbs sampling iterations $N_{\text{iter}}$, measurements $\mathbf{z}_k$, 
\Ensure
Approximate PMBM filtering density $f_{k|k}(\widetilde{\mathbf{x}}_k)$.
\State Initialize the blocked Gibbs sampler, see Sec. \ref{sec_initialization}.
\For{$i = 1$ to $N_{\text{iter}}$}
    \State Compute updated global and local hypotheses using the sampled previous global hypothesis and measurement associations from last sampling iteration, see Proposition \ref{pmbm_update}.
    \State Sample a set of detected object states using \eqref{eq_set_detected_conditional} for the full Gibbs sampler, or using \eqref{eq_set_of_all_samples} for the collapsed Gibbs sampler.
    \State Sample a previous global hypothesis using \eqref{eq_sample_a} for the full Gibbs sampler and \eqref{eq_collapsed_sample_a} for the collapsed Gibbs sampler.
    \State Sample measurement-oriented  associations using \eqref{eq_sample_beta1}-\eqref{eq_sample_beta2} for the full Gibbs sampler and \eqref{eq_collapsed_sample_beta1}-\eqref{eq_collapsed_sample_beta2} for the collapsed Gibbs sampler.
\EndFor
\State Compute updated Poisson intensity \eqref{eq_likelihood_missed_ppp} for undetected objects.
\State Global and local hypothesis reduction.
\end{algorithmic}
\end{algorithm}

The pseudocode for the PMBM update using blocked Gibbs samplers (both full and collapsed) is provided in Algorithm \ref{alg_full_gibbs}. By running the Gibbs sampler for $N_{\text{iter}}$ iterations, we obtain $N_{\text{iter}}$ global hypotheses with repetition, and only unique global hypotheses are kept.

As detailed in Appendix~\ref{app:complexity}, the computational complexity of one complete sampling iteration is
\begin{equation}
    \mathcal{O}\left(
    |\mathcal{A}_{k|k-1}|n_{k|k-1}
    +m_k(n_{k|k-1}+m_k)
    \right).
    \label{eq_gibbs_complexity}
\end{equation}
Thus, conditioned on the predicted global hypothesis set $\mathcal{A}_{k|k-1}$, the per-iteration complexity is quadratic in the number of measurements $m_k$ and linear in the number of existing Bernoulli components $n_{k|k-1}$. The complexity is also linear in $|\mathcal{A}_{k|k-1}|$, whose size is capped at $N_{\text{iter}}$ and also controlled through global hypothesis pruning.

The following proposition establishes asymptotic convergence of the full blocked Gibbs sampler, but the collapsed sampler does not generally inherit this guarantee because of Approximations A1 and A2. Nevertheless, the result in Section \ref{sec_simulation} shows empirically that increasing the sampling budget improves performance, with diminishing returns at larger iteration counts.

\begin{proposition}
Suppose that all retained previous global hypotheses have positive weights and that the single object densities, Poisson intensities, and single object measurement likelihoods appearing in the exact full conditional distributions \eqref{eq_sample_a}, \eqref{eq_sample_beta}, and \eqref{eq_set_detected_conditional} are all positive on their feasible supports. 
Suppose further that $\lambda_k^C(z_k^j)>0$ for all $j\in\{1,\dots,m_k\}$. Then the full blocked Gibbs sampler in Sec. \ref{sec_full_Gibbs} is irreducible and aperiodic on the feasible support of the joint posterior
${f}_{k|k}^d(\widetilde{\mathbf{x}}_k^d,a,\beta_k)$
in \eqref{eq_joint_mbm_factorization}. This posterior is also the unique stationary distribution of the Gibbs sampler, and the sample distribution converges to it in total variation distance for almost every feasible initialization \cite{tierney1994markov}.
\label{prop:gibbs_convergence}
\end{proposition}

Proposition \ref{prop:gibbs_convergence} is proved in Appendix \ref{app:gibbs_convergence}. Note that it requires the condition $\lambda_k^C(z_k^j)>0$ for all $j\in\{1,\dots,m_k\}$, which allows a new Bernoulli component associated with a single measurement to become non-existent, such that its initiating measurement can be regarded as clutter and subsequently reassigned to other Bernoulli components. If $\lambda_k^C(z_k^j)=0$, then the existence probability of the $j$-th new Bernoulli component is one, and the separate local hypothesis and association updates might make the chain reducible. The zero-clutter case can be handled by jointly sampling the existence of the new Bernoulli component and the association of its initiating measurement. Such a degenerate case is not considered in this paper.

\subsection{Track-Oriented Implementation}\label{sec_track_oriented}

The conditional distributions of previous global hypotheses $a$ and measurement associations $\beta_k$ have simple expressions that allow for efficient sampling. However, sampling the set $\widetilde{\mathbf{x}}_k^d$ of detected objects (with and without collapsing the object existence variables) from \eqref{eq_set_detected_conditional} requires first computing the updated MB densities and is thus more computationally demanding. In addition, in the iterative sampling process, the same set of measurements could be associated with the same Bernoulli component multiple times, resulting in the redundant computation of the same updated local hypothesis densities. To address this inefficiency, we employ a track-oriented implementation of the extended object PMBM filter as in \cite{xia2019extended}, where the local hypotheses of each Bernoulli component have a tree structure, and global hypotheses are encoded implicitly through a look-up table. In addition, we store the local measurement associations of updated local hypotheses in a hash table. Together, these ensure that each updated local hypothesis only needs to be computed once. 

The sampling of previous global hypotheses can also be carried out efficiently. In particular, the likelihood of a previous global hypothesis decomposes into a product of factors associated with individual existing Bernoulli components. These factors can be precomputed and stored since each of them depends only on the corresponding local hypothesis Moreover, sampling from the resulting categorical distribution depends on relative likelihood values across previous global hypotheses, which may share common local hypotheses. Therefore, any multiplicative term shared by all global hypotheses can be omitted from the computation.

The global hypothesis weights can be computed and stored when constructing the updated MB densities \eqref{eq_set_detected_conditional}. Let $\widehat{\mathcal{A}}_{k|k}\subseteq\mathcal{A}_{k|k}$ denote the set of unique updated global hypotheses discovered during the sampling process. For each $a\in\widehat{\mathcal{A}}_{k|k}$, its weight $w^a_{k|k}$ is computed, independently of how many times the global hypothesis is visited. Duplicate samples therefore do not increase the weight of a global hypothesis. The resulting MBM is thus restricted to the sampled global hypothesis set and the MB weights are renormalized over that set.

We note that this interpretation differs from a conventional frequency-based MCMC approximation, in which posterior probabilities are estimated from empirical sampling frequencies after convergence. Here, the blocked Gibbs sampler instead acts as a stochastic search mechanism that determines which global hypotheses are included, while their weights are determined by the PMBM update. Consequently, samples from the early iterations need not be discarded since every unique sampled hypothesis can be retained and assigned its weight. Nevertheless, sufficient exploration remains important, because a high-weight global hypothesis that is never visited cannot be included in the truncated MBM approximation.

To further reduce computational complexity, we use standard reduction strategies for PMBM densities. These include discarding global hypotheses with small weights, removing Bernoulli components with low existence probabilities or those not supported by the remaining global hypotheses, and simplifying the Poisson intensity. Details of these procedures can be found in \cite{pmbmextended2,xia2019extended}.

\begin{remark}
We clarify the relations between the proposed extended object PMBM filters based on blocked Gibbs sampling and the AbNHPP tracker in \cite{li2023adaptive}, which also relies on the PPP measurement model and Gibbs sampling for multi-object inference, but assumes a fixed and known number of objects. From a high-level perspective, the AbNHPP tracker can be interpreted as a special case of the proposed PMBM filter. In particular, if the multi-object birth and death processes are removed from the PMBM recursion and an initial MB(M) prior is assumed at time step $0$, then the PMBM density reduces to an MBM density with all Bernoulli components having existence probability one. A further difference lies in the posterior representation: the AbNHPP tracker in \cite{li2023adaptive} adopts a Rao--Blackwellized particle representation, in which empirical sampling frequencies are used and burn-in samples must therefore be discarded. In contrast, the proposed PMBM filters employ Gibbs sampling to discover global hypotheses and assign their weights using the PMBM update equations.
\end{remark}

\begin{remark}
To extend the proposed algorithms to the standard measurement model for extended objects, consisting of a zero-inflated PPP model \cite{phdextended2}, a binary detection variable $d_k^i$ can be introduced for each existing Bernoulli component. Specifically, conditioned on $d_k^i=1$, the object generates a Poisson set of measurements according to \eqref{eq_ppp_meas}, whereas $d_k^i=0$ means missed detection. The joint posterior \eqref{eq_joint_mbm_factorization} can then be augmented with these detection variables to facilitate factorization. Each iteration of the exact blocked Gibbs sampler would additionally sample them, and they may be marginalized using suitable approximations in the collapsed sampler. A detailed treatment of this extension is left for future work.
\end{remark}

\section{Simulations and Results}
\label{sec_simulation}

\subsection{GGIW Implementation and Simulation Setup}
\label{sec_simulation_setup}

We use the random matrix model for EOT, in which each object has an elliptical shape \cite{randomMatrix2}. The single object state is written as $x_k=(\gamma_k,\xi_k,X_k)$, and the measurement likelihood is modeled as Gaussian:
\begin{equation}
  \ell_k(z | x_k) = \mathcal{N}\left(z;H\xi_k,X_k\right). \label{eq_single_measurement_likelihood}
\end{equation}
Here, $\gamma_k>0$ denotes the Poisson measurement rate, $\xi_k\in\mathbb{R}^{n_x}$ is the kinematic state, and $X_k$ is a symmetric positive definite random matrix describing the object extent. The matrix $H$ maps the kinematic state $\xi_k$ to the measurement space $\mathbb{R}^{n_z}$, and we assume that the measurement noise is negligible as compared to the object extent.

For the PPP measurement model in \eqref{eq_ppp_meas} with the single measurement likelihood in \eqref{eq_single_measurement_likelihood}, the single-object density is represented by a GGIW distribution \cite{randomMatrix2,phdextended}
\begin{align}
  f_{k|k}(x_k) &= \mathcal{G}(\gamma_k;\alpha_{k|k},\beta_{k|k})\mathcal{N}(\xi_k;\bar{x}_{k|k},P_{k|k})\nonumber \\
  &~~~\times \mathcal{IW}(X_k;v_{k|k},V_{k|k}), \label{eq_ggiw}
\end{align}
where $\mathcal{G}(\cdot;\alpha_{k|k},\beta_{k|k})$ denotes a Gamma distribution with shape parameter $\alpha_{k|k}>0$ and inverse-scale parameter $\beta_{k|k}>0$, whereas $\mathcal{IW}(\cdot;v_{k|k},V_{k|k})$ denotes an inverse-Wishart distribution with degree of freedom $v_{k|k}>d+1$ and parameter matrix $V_{k|k}$\footnote{We use the inverse-Wishart parameterization in \cite{gelman1995bayesian}, under which the mean of $\mathcal{IW}(X_k;v_{k|k},V_{k|k})$ is $\hat{X}_{k|k}=V_{k|k}/(v_{k|k}-d-1)$.}.

The GGIW prediction and update equations, the corresponding predicted likelihoods, and the Poisson intensity representation are provided in Appendix~\ref{app:ggiw_implementation}. The proposed samplers draw the Gamma and Gaussian variables directly and obtain inverse-Wishart samples by first sampling a Wishart matrix and then taking its inverse \cite{gelman1995bayesian}.

We assess the proposed extended object PMBM filter implementations using blocked Gibbs sampling through Monte Carlo simulations. We employ the same simulation setup as in \cite{florian2021scalable,xia2023trajectory,xia2023efficient}, with the region of interest defined by $[-150\,\text{m}, 150\,\text{m}] \times [-150\,\text{m}, 150\,\text{m}]$. Ten objects are initialized at positions drawn uniformly on a circle of radius $125\,\text{m}$ centered at the origin, each with an initial speed directed toward the center. As they move inward, the ten objects become closely spaced for a period of time and then separate.

The proposed two extended object PMBM filters using blocked Gibbs sampling are evaluated against the PMBM filter using C\&A \cite{pmbmextended2}, the sampling-based PMBM filter using stochastic optimization \cite{soextended} and the PMB filter with particle BP \cite{xia2023trajectory} across different scenarios and parameter settings. We consider two initialization strategies for both the full and collapsed blocked Gibbs samplers. The first is a simple initialization that first selects the previous global hypothesis with the largest weight and then assigns the $j$-th measurement to the $j$-th new Bernoulli component for each $j\in\{1,\dots,m_k\}$. The second follows the initialization procedure described in Section~\ref{sec_initialization}. The compared implementations are as follows:
\begin{itemize}
  \item Extended object PMBM filter using blocked Gibbs sampling, referred to as PMBM-Gibbs1 and PMBM-Gibbs2 for the first and second initialization methods, respectively\footnote{MATLAB code will be available at \url{https://github.com/yuhsuansia} upon acceptance.}.
  \item Extended object PMBM filter with collapsed blocked Gibbs sampling, referred to as PMBM-C-Gibbs1 and PMBM-C-Gibbs2 for the first and second initialization methods, respectively\footnotemark[\value{footnote}].
  \item Extended object PMB filter with particle BP, referred to as PMB-BP\footnote{The implementation is based on the MATLAB code of \cite{xia2023trajectory}, available at \url{https://github.com/yuhsuansia/Trajectory-PMB-EOT-BP}, which has been modified to account for an unknown Poisson measurement rate.}.
  \item Extended object PMBM filter using C\&A, referred to as PMBM-CA\footnote{MATLAB implementations of PMBM-CA and PMBM-SO are available at \url{https://github.com/yuhsuansia/Extended-target-PMBM-tracker}.}.
  \item Extended object PMBM filter using stochastic optimization, referred to as PMBM-SO\footnotemark[\value{footnote}].
\end{itemize}

\begin{figure}[!t]
    \centering
    \includegraphics[width=\columnwidth]{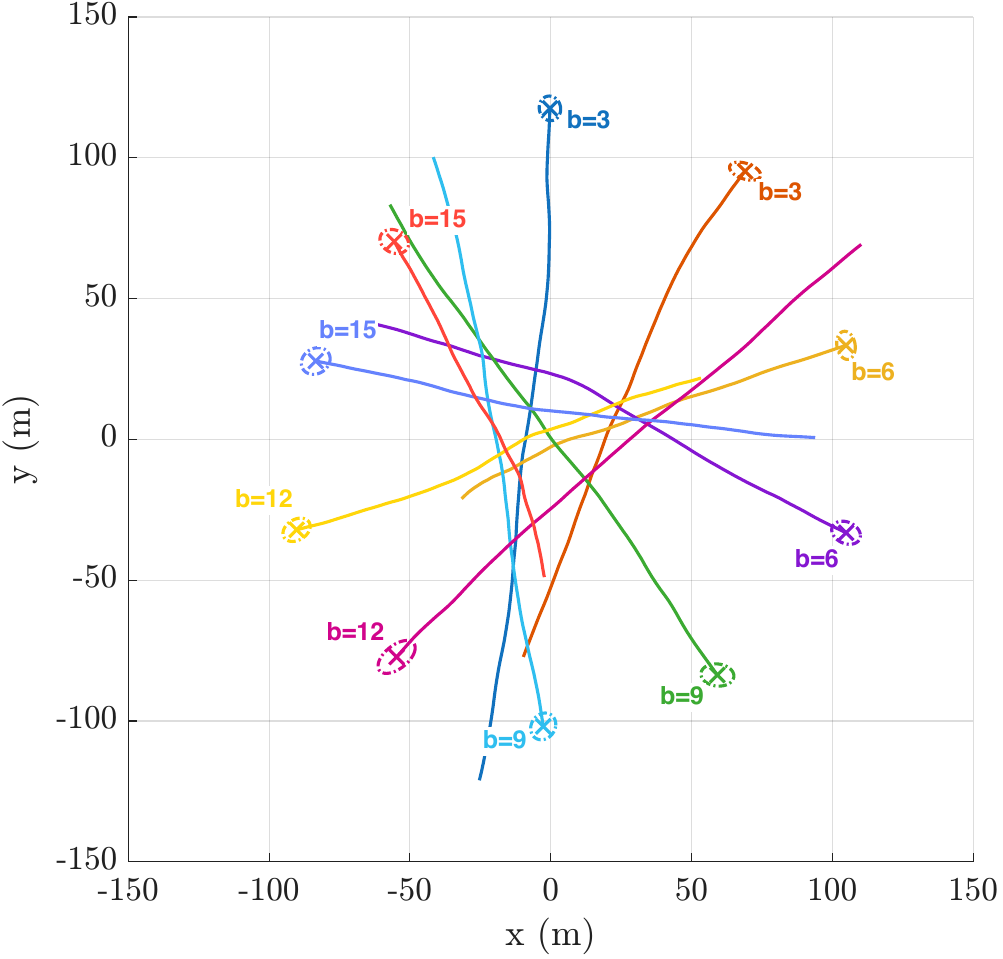}
    \caption{Illustration of the ground truth scenario. Two objects are born at each of the time steps 3, 6, 9, 12, and 15, while two objects are dead at time steps 83, 86, 89, 92, and 95. The initial object locations are indicated by crosses, along with their corresponding elliptical extents. Text annotations specify the birth times; e.g., $b = k$ denotes that an object appears at time step $k$.}
    \label{fig_scenario}
\end{figure}

The object kinematic state is  $\xi_k=[p_x,v_x,p_y,v_y]^T$, containing 2D position and velocity. Each object moves according to a nearly constant velocity model, with transition matrix and process noise covariance
\begin{equation}
  F = I_2 \otimes \begin{bmatrix}
    1 & T_s \\
    0 & 1
  \end{bmatrix},\quad Q = \sigma_q^2 I_2 \otimes \begin{bmatrix}
    T_s^3/3 & T_s^2/2 \\
    T_s^2/2 & T_s
  \end{bmatrix},
\end{equation}
where $I_2$ is an $2\times2$ identity matrix, $\otimes$ is the Kronecker product, $T_s=0.2\,\text{s}$ is the sampling interval, and $\sigma_q=0.8$. For each object, the initial extent matrix is drawn from an inverse-Wishart distribution with degree of freedom $v_0=100$ and mean $\mathrm{diag}([5,5])$, while the initial Poisson measurement rate is drawn from a Gamma distribution with shape  $\alpha_0=100$ and mean $\gamma$. Both the object extent and the measurement rate are assumed constant over time. The true object trajectories are shown in Fig.~\ref{fig_scenario}. We use an object survival probability of $p^S=0.99$, and set $\eta=1.01$ and $\tau=100T_s$ in \eqref{eq_ggiw_motion}.

For all implementations, the Poisson birth rate is set to $0.01$. In the birth model, the velocity follows a zero-mean Gaussian distribution with covariance $225I_2$, the extent follows an inverse-Wishart distribution with mean $5I_2$ and degree of freedom $4$, and the measurement rate follows a Gamma distribution with mean $\gamma_0$ and rate $100$. For all the GGIW-based implementations, the object position in the birth density is modeled by a zero-mean Gaussian with covariance $150^2 I_2$, whereas for PMB-BP it is taken to be uniformly distributed over the entire region of interest.

For the measurement model, the observation matrix is $H = I_2 \otimes [1~0]$, and the clutter is uniformly distributed over the entire region of interest, with a Poisson rate $\gamma^C$. In the simulations, three different configurations are considered, corresponding to $(\gamma,\gamma^C) \in \{(5,10), (10,10), (5,20)\}$.

PMBM-CA employs DBSCAN to partition the measurement set into clusters. To obtain suitable measurement partitions, a grid search is performed over the DBSCAN clustering threshold, with the parameter varying from 0.1 to 5 in increments of 0.2. For each distinct measurement partition and each previous global hypothesis with weight $w^a_{k|k-1}$, Murty's algorithm is then employed to generate the $\lceil 100 w^a_{k|k-1}\rceil$-best global hypotheses. PMBM-SO uses stochastic optimization to iteratively sample data associations, and the sampling iteration is set to $\lceil 30n_{k|k} w^a_{k|k-1}\rceil$. In addition, the initialization procedure described in Section~\ref{sec_initialization} is also adopted, with the same DBSCAN clustering threshold $\sqrt{5}$ and minimum cluster size $1$. For PMBM-based implementations, we prune global hypotheses with weights smaller than $10^{-3}$ and  Bernoulli components with existence probabilities below $10^{-3}$. We also prune GGIW components in the Poisson intensity of undetected objects with weight smaller than $10^{-3}$.

PMB-BP uses a particle representation, in which each particle consists of the kinematic state, the extent matrix, and the Poisson object measurement rate. New particles are generated from a proposal distribution composed of  Gaussian transition density for the kinematic state with mean-preserving Wishart and Gamma distributions for the extent matrix and Poisson measurement rate, respectively, using small uncertainties that are selected to match the GGIW prediction parameters in \eqref{eq_ggiw_motion}. Systematic resampling is performed at every time step. For BP, the number of message passing iterations is set to 3, and we adopt the measurement-driven initialization strategy for newly detected objects from \cite[Sec. VI]{xia2023trajectory}. We also apply message censoring and measurement reordering to support new track initialization \cite{meyer2020scalable2}. In addition, Bernoulli components with  probability of existence below $10^{-3}$ are pruned.

To obtain state estimates from the filtering density, the PMBM filters first select the global hypothesis with the largest weight and then use the mean of each single-object density associated with Bernoulli components whose existence probabilities are greater than $0.5$. For PMB-BP, the state estimate of each Bernoulli component with existence probability above $0.5$ is computed as the weighted average of its particles. Multi-object filtering performance is evaluated using the GOSPA metric \cite{gospa}, with the square-root Gaussian Wasserstein distance as the base measure for comparing  extended object states with elliptical shape \cite{gwd}. In the GOSPA evaluation, we set $\alpha=2$, the cutoff distance to $c=20$, and the exponent to $p=1$. All results are obtained by averaging over 100 Monte Carlo runs.

\subsection{Results and Discussion}
\label{sec_results}

\begin{table*}[!t]
  \caption{Multi-object filtering performance evaluated using the GOSPA metric, together with its decomposition into state estimation, missed detection, and false detection errors, averaged over 100 time steps.}
  \resizebox{\textwidth}{!}{%
\begin{tabular}{llcccccccccccc}
\toprule
 &  & \multicolumn{4}{c}{$\gamma = 5$, $\lambda^C = 10$} & \multicolumn{4}{c}{$\gamma = 10$, $\lambda^C = 10$} & \multicolumn{4}{c}{$\gamma = 5$, $\lambda^C = 20$} \\
\cmidrule(lr){3-6}\cmidrule(lr){7-10}\cmidrule(lr){11-14}
 &  & Total & State & Miss & False & Total & State & Miss & False & Total & State & Miss & False \\
\midrule
PMBM-Gibbs1 & \# iterations = 500 & 26.91 & 11.42 & 13.06 & 2.44 & 68.73 & 11.75 & 52.65 & 4.33 & 40.25 & 11.69 & 24.10 & 4.46 \\
\midrule
\multirow{3}{*}{PMBM-Gibbs2}
& \# iterations = 20 & 10.10 & 8.44 & 1.12 & 0.54 & 9.62 & 7.71 & 0.11 & 1.80 & 10.34 & 8.42 & 1.15 & 0.76 \\
& \# iterations = 100 & 9.62 & 8.08 & 1.47 & 0.07 & 8.47 & 7.45 & 0.36 & 0.67 & 10.03 & 8.20 & 1.64 & 0.19 \\
& \# iterations = 500 & 9.49 & 8.07 & 1.19 & 0.23 & 8.04 & 7.42 & 0.10 & 0.51 & 9.75 & 8.20 & 1.39 & 0.16 \\
\midrule
\multirow{3}{*}{PMBM-C-Gibbs1}
& \# iterations = 20 & 10.10 & 8.43 & 1.20 & 0.47 & 9.16 & 7.80 & 0.23 & 1.13 & 10.56 & 8.35 & 1.69 & 0.51 \\
& \# iterations = 100 & 9.39 & 8.17 & 1.04 & 0.18 & 8.37 & 7.38 & 0.66 & 0.33 & 9.71 & 8.29 & 1.23 & 0.20 \\
& \# iterations = 500 & 9.06 & 7.89 & 0.97 & 0.20 & 8.09 & 7.48 & 0.39 & 0.22 & 9.10 & 8.01 & 1.02 & 0.07 \\
\midrule
\multirow{3}{*}{PMBM-C-Gibbs2}
& \# iterations = 20 & 9.90 & 8.36 & 0.95 & 0.59 & 9.08 & 7.41 & 1.33 & 0.33 & 10.31 & 8.38 & 1.00 & 0.92 \\
& \# iterations = 100 & 9.22 & 8.06 & 0.93 & 0.23 & 8.20 & 7.41 & 0.56 & 0.23 & 9.28 & 8.16 & 0.97 & 0.15 \\
& \# iterations = 500 & \textbf{8.79} & 7.87 & 0.85 & 0.07 & 8.00 & 7.39 & 0.41 & 0.21 & \textbf{8.88} & 7.87 & 0.90 & 0.11 \\
\midrule
\multirow{3}{*}{PMB-BP}
& \# particles = 1000 & 14.67 & 13.03 & 0.73 & 0.91 & 15.45 & 10.35 & 0.05 & 5.06 & 15.41 & 13.35 & 0.76 & 1.29 \\
& \# particles = 5000 & 10.27 & 9.48 & 0.61 & 0.17 & 9.04 & 7.64 & 0.02 & 1.37 & 10.60 & 9.60 & 0.66 & 0.34 \\
& \# particles = 10000 & 9.16 & 8.47 & 0.60 & 0.09 & \textbf{7.71} & 6.84 & 0.02 & 0.85 & 9.49 & 8.65 & 0.63 & 0.21 \\
\midrule
PMBM-CA & -- & 16.87 & 9.95 & 4.02 & 2.90 & 16.03 & 8.85 & 6.14 & 1.04 & 17.02 & 9.99 & 4.19 & 2.84 \\
PMBM-SO & -- & 10.02 & 8.13 & 1.11 & 0.79 & 8.81 & 7.24 & 0.82 & 0.74 & 10.03 & 8.13 & 1.05 & 0.84 \\
\bottomrule
\end{tabular}
  }
  \label{tab_gospa}
\end{table*}

\begin{table}[!t]
  \caption{Runtime per Monte Carlo run (in seconds)}
  \resizebox{\columnwidth}{!}{%
\begin{tabular}{llccc}
\toprule
 &  & \makecell{$\gamma = 5$, \\$\lambda^C = 10$} & \makecell{$\gamma = 10$, \\$\lambda^C = 10$} & \makecell{$\gamma = 5$, \\$\lambda^C = 20$} \\
\midrule
PMBM-Gibbs1 & \# iterations = 500 & 93.5 & 116.5 & 109.8 \\
\midrule
\multirow{3}{*}{PMBM-Gibbs2}
& \# iterations = 20 & 5.2 & 6.4 & 6.2 \\
& \# iterations = 100 & 19.8 & 25.4 & 23.4 \\
& \# iterations = 500 & 89.6 & 123.0 & 110.8 \\
\midrule
\multirow{3}{*}{PMBM-C-Gibbs1}
& \# iterations = 20 & 5.7 & 7.0 & 6.8 \\
& \# iterations = 100 & 21.9 & 28.2 & 27.7 \\
& \# iterations = 500 & 105.1 & 138.2 & 128.6 \\
\midrule
\multirow{3}{*}{PMBM-C-Gibbs2}
& \# iterations = 20 & 5.0 & 6.7 & 6.9 \\
& \# iterations = 100 & 21.9 & 27.9 & 27.6 \\
& \# iterations = 500 & 103.5 & 140.7 & 130.9 \\
\midrule
\multirow{3}{*}{PMB-BP}
& \# particles = 1000 & 12.8 & 23.3 & 13.5 \\
& \# particles = 5000 & 62.4 & 117.8 & 70.2 \\
& \# particles = 10000 & 132.8 & 224.9 & 155.4 \\
\midrule
PMBM-CA & -- & 230.3 & 250.2 & 227.0 \\
PMBM-SO & -- & 53.9 & 125.4 & 74.8 \\
\bottomrule
\end{tabular}
  }
  \label{tab_runtime}
\end{table}

\begin{figure*}[t]
\centering

\subfloat{\includegraphics[width=0.24\textwidth]{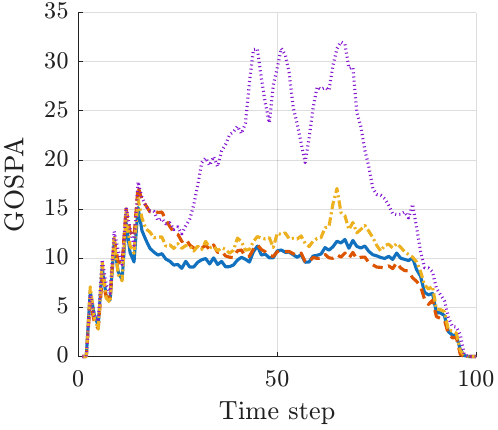}}
\subfloat{\includegraphics[width=0.24\textwidth]{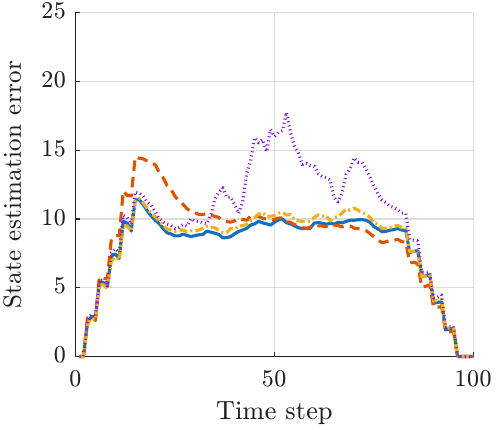}}
\subfloat{\includegraphics[width=0.24\textwidth]{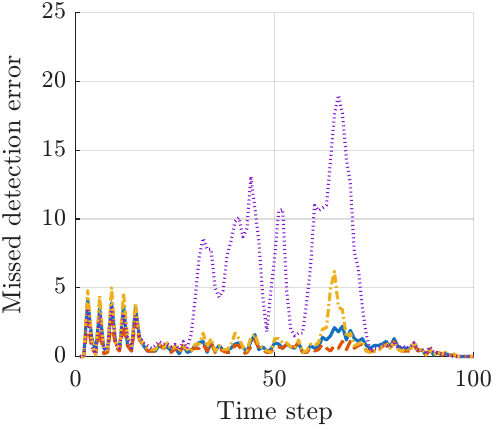}}
\subfloat{\includegraphics[width=0.24\textwidth]{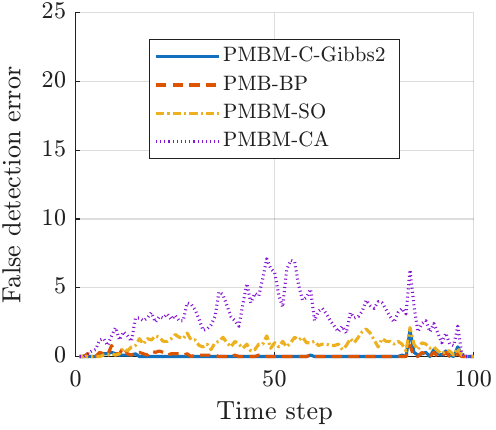}}

\par\smallskip
{\footnotesize (a) $\gamma = 5$, $\lambda^C = 10$}

\subfloat{\includegraphics[width=0.24\textwidth]{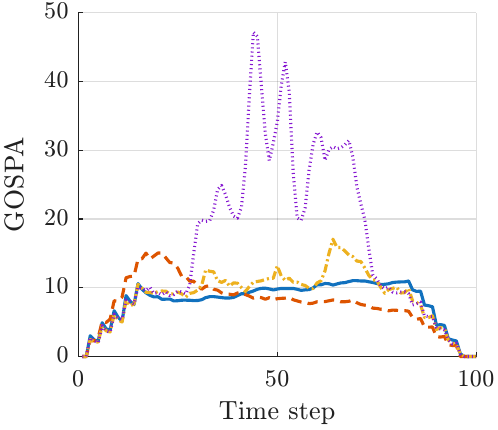}}
\subfloat{\includegraphics[width=0.24\textwidth]{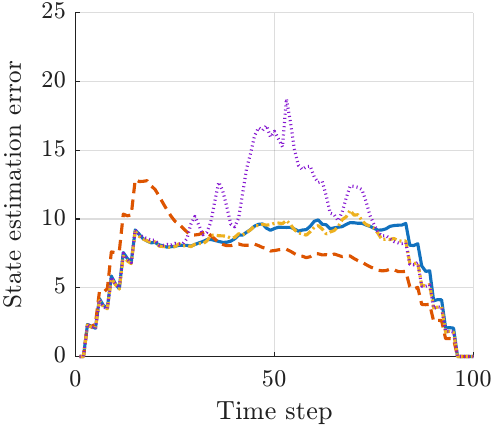}}
\subfloat{\includegraphics[width=0.24\textwidth]{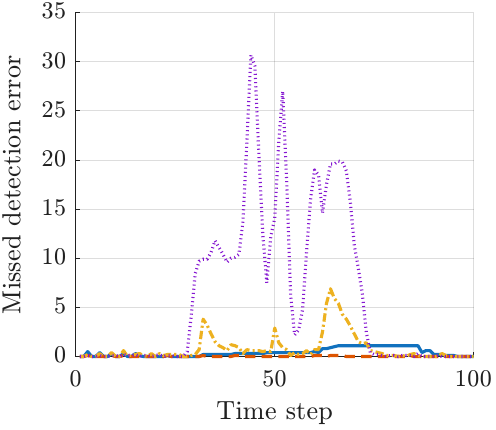}}
\subfloat{\includegraphics[width=0.24\textwidth]{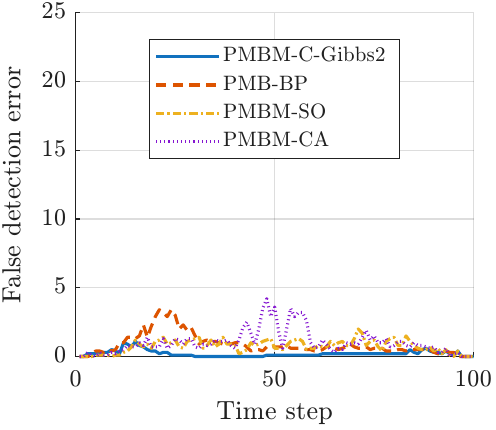}}

\par\smallskip
{\footnotesize (b) $\gamma = 10$, $\lambda^C = 10$}

\subfloat{\includegraphics[width=0.24\textwidth]{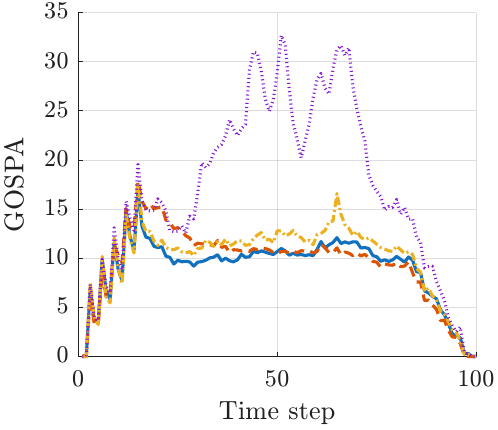}}
\subfloat{\includegraphics[width=0.24\textwidth]{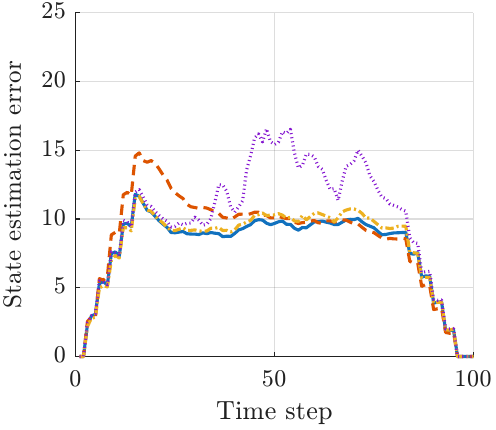}}
\subfloat{\includegraphics[width=0.24\textwidth]{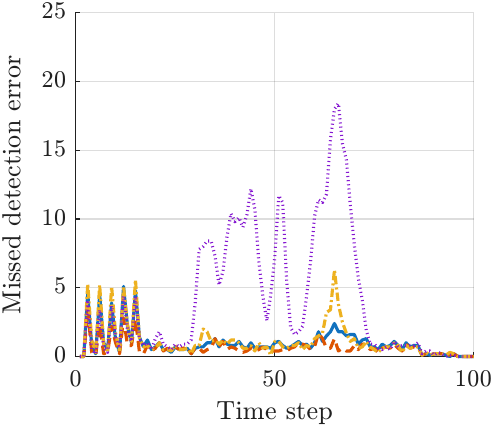}}
\subfloat{\includegraphics[width=0.24\textwidth]{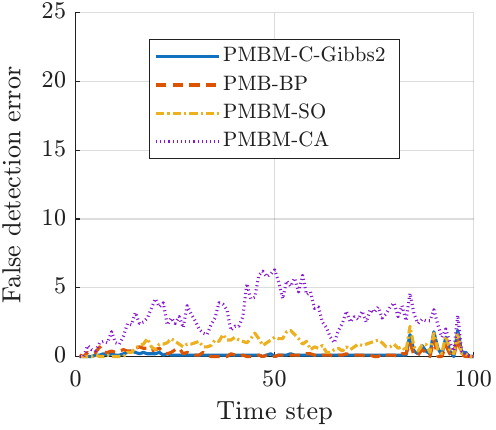}}

\par\smallskip
{\footnotesize (c) $\gamma = 5$, $\lambda^C = 20$}

\caption{Performance comparison of PMBM-C-Gibbs2 (with 500 sampling iterations), PMB-BP (with 10000 particles), PMBM-SO and PMBM-CA, showing the GOSPA error and its decomposition over time for three different scenarios.}
\label{fig:results}
\end{figure*}

The PMBM implementations based on blocked Gibbs sampling with 20, 100, and 500 Gibbs sampling iterations are compared with PMBM-CA, PMBM-SO, and PMB-BP using 1000, 5000, and 10000 particles. Table~\ref{tab_gospa} reports the estimation performance, measured by GOSPA error and its decomposition into state estimation, missed detection, and false detection errors, averaged over 100 time steps. As expected, increasing the number of sampling iterations or particles consistently improves performance for PMBM using blocked Gibbs sampling and PMB-BP. For PMBM implementations using blocked Gibbs sampling, the improvement from 100 to 500 iterations is generally smaller than that from 20 to 100 iterations, indicating diminishing returns as the sampling budget increases. This provides empirical evidence of the practical convergence behavior of the proposed samplers. Moreover, all methods perform better in the second scenario, which has a higher object measurement rate, and worse in the third scenario, which has a higher clutter rate, relative to the first scenario.

It can be seen from the results in Table \ref{tab_gospa} and Table \ref{tab_runtime} that all PMBM variants based on blocked Gibbs sampling, except PMBM-Gibbs1, outperform PMBM-CA and PMBM-SO in terms of both GOSPA error and computational runtime. In addition, PMBM-C-Gibbs2 performs best in the first and third scenarios with $\gamma=5$, whereas PMB-BP has the best performance in the second scenario with $\gamma=10$. These results indicate that the proposed blocked Gibbs sampling method provides a more effective approach to data association than C\&A and stochastic optimization for multiple EOT under the considered settings, while remaining competitive with PMB-BP.

A closer comparison among the PMBM variants with blocked Gibbs sampling further reveals the importance of initialization and collapsing. PMBM-Gibbs1 has difficulty detecting newborn objects, particularly in the second scenario, where the higher Poisson object measurement rate leads to larger missed detection errors. The clustering-based initialization used in PMBM-Gibbs2 substantially improves the detection of newborn objects, whereas the corresponding improvement for PMBM-C-Gibbs2 is comparatively modest. This indicates that the collapsed Gibbs sampler is much less sensitive to initialization and can still achieve strong estimation performance with a relatively simple initialization scheme. The results are also consistent with the interpretation that collapsing facilitates the exploration of high-weight global hypotheses. We note that this advantage is obtained even though the sampling distribution of PMBM-C-Gibbs2 is not the exact posterior over global hypotheses.

To examine these performance differences in greater detail, Fig.~\ref{fig:results} presents the time evolution of the GOSPA error and its decomposition for PMBM-C-Gibbs2 with 500 Gibbs sampling iterations, PMBM-CA, PMBM-SO, and PMB-BP with 10,000 particles across the three scenarios. PMBM-CA exhibits substantial missed detection errors when multiple objects move in close proximity. Although PMBM-SO alleviates this issue, its missed detection error remains elevated when closely spaced objects are about to separate. PMBM-C-Gibbs2 substantially reduces these errors, showing the advantage of blocked Gibbs sampling in resolving highly uncertain data associations.

We also observe that PMBM-C-Gibbs2 generally incurs lower false detection error, except immediately after objects disappear, whereas PMB-BP consistently exhibits lower missed detection error, both when new objects are born and after previously closely spaced objects become separated. This difference primarily stems from the birth initialization strategy. In GGIW-PMBM implementations, the birth density is represented by Gaussian components with large covariances to cover the region of interest. In contrast, PMB-BP initializes newly detected objects using Gaussian proposal distributions with small covariances centered at the mean locations of measurement clusters. This enables PMB-BP to establish new tracks more readily, but also increases its susceptibility to false estimates, particularly in the second scenario with a higher Poisson object measurement rate.

Another contributing factor is that, when objects first move into close proximity and later separate, the particle-based representation in PMB-BP can better capture the multimodal structure of the posterior, which further helps reduce missed detection error. We also observe that the PMBM-based methods generally achieve lower state estimation error than PMB-BP before the objects become closely spaced. However, in the second scenario with a higher Poisson object measurement rate, PMB-BP with 10,000 particles attains the lowest state estimation error during the period when the objects become closely spaced and subsequently separate. One possible explanation is that the analytical GGIW update in \eqref{eq_ggiw_update_paras} is only approximate. Therefore, when more measurements and particles are available, PMB-BP may yield more accurate object state estimates.

Table~\ref{tab_runtime} reports the average runtime per Monte Carlo run in seconds for the different implementations\footnote{MATLAB implementations on Intel Xeon ICX Platinum 8358 2.6GHz under the same computational resource setting. No explicit parallel computation was used; in particular, the measurement association variables in the proposed samplers were processed serially.}. The results show that PMBM-Gibbs1 and PMBM-C-Gibbs2 have similar runtime, while the clustering-based initialization leads to a modest runtime increase. Combined with the GOSPA errors in Table~\ref{tab_gospa}, we observe that PMBM-C-Gibbs2 with only 20 sampling iterations already performs on par with PMB-BP using 5000 particles, while being roughly an order of magnitude faster. Overall, the results show that the proposed PMBM filter using blocked Gibbs sampling offers a more favorable trade-off between estimation accuracy and runtime.

\section{Conclusions}

This paper has presented efficient implementations of the extended object PMBM filter that propagate global hypotheses with non-negligible weights via blocked Gibbs sampling. We have further introduced a collapsed Gibbs sampling strategy, which marginalizes out the Bernoulli object existence variables, leading to increased robustness to initialization and fast exploration of high-weight global hypotheses. The proposed extended object PMBM filters have been evaluated in a challenging scenario involving ten objects moving in proximity and compared with two PMBM filters using C\&A and stochastic optimization and a PMB filter based on particle BP. The results demonstrate that the proposed PMBM filters achieve a more favorable trade-off between estimation accuracy and runtime.

Future research directions include extending the proposed extended object PMBM filters to sets of trajectories to preserve full trajectory information, as in \cite{xia2019extended,xia2023trajectory,xia2023efficient}, incorporating an MB birth model, adopting a zero-inflated Poisson measurement model to account for object detection probability, and investigating more expressive extended object models. Another promising direction is the development of more efficient collapsed Gibbs sampling schemes, e.g., by marginalizing out the object states, either partially or fully, in addition to the object existence.

\bibliographystyle{IEEEtran}
\bibliography{mybibli.bib}

@article{extendedoverview,
  title={Extended object tracking: {I}ntroduction, overview and applications},
  author={Granstr{\"o}m, Karl and Baum, Marcus and Reuter, Stephan},
  journal={Journal of Advances in Information Fusion},
  volume={12},
  number={2},
  year={2017}
}

@book{rfs,
  title={Advances in Statistical Multisource-Multitarget Information Fusion},
  author={Mahler, Ronald PS},
  year={2014},
  publisher={Artech House Norwood, MA}
}

@article{ppp,
  title={Spatial distribution model for tracking extended objects},
  author={Gilholm, Kevin and Salmond, David},
  journal={IEE Proceedings-Radar, Sonar and Navigation},
  volume={152},
  number={5},
  pages={364--371},
  year={2005},
  publisher={IET}
}

@article{meyer2020scalable,
  title={Scalable Data Association for Extended Object Tracking},
  author={Meyer, Florian and Win, Moe Z},
  journal={IEEE Transactions on Signal and Information Processing over Networks},
  volume={6},
  pages={491--507},
  year={2020},
  publisher={IEEE}
}

@article{meyer2018message,
  title={Message passing algorithms for scalable multitarget tracking},
  author={Meyer, Florian and Kropfreiter, Thomas and Williams, Jason L and Lau, Roslyn and Hlawatsch, Franz and Braca, Paolo and Win, Moe Z},
  journal={Proceedings of the IEEE},
  volume={106},
  number={2},
  pages={221--259},
  year={2018},
  publisher={IEEE}
}

@inproceedings{meyer2020scalable2,
  title={Scalable Detection and Tracking of Extended Objects},
  author={Meyer, Florian and Williams, Jason L},
  booktitle={IEEE International Conference on Acoustics, Speech and Signal Processing (ICASSP)},
  pages={8916--8920},
  year={2020},
  organization={IEEE}
}

@inproceedings{yang2018linear,
  title={Linear-time joint probabilistic data association for multiple extended object tracking},
  author={Yang, Shishan and Thormann, Kolja and Baum, Marcus},
  booktitle={10th Sensor Array and Multichannel Signal Processing Workshop},
  pages={6--10},
  year={2018},
  organization={IEEE}
}

@inproceedings{garcia2019gaussian,
  title={Gaussian implementation of the multi-{B}ernoulli mixture filter},
  author={Garc{\'i}a-Fern{\'a}ndez, {\'A}ngel F and Xia, Yuxuan and Granstr{\"o}m, Karl and Svensson, Lennart and Williams, Jason L},
  booktitle={22th International Conference on Information Fusion},
  pages={1--8},
  year={2019},
  organization={IEEE}
}

@article{randomMatrix2,
  title={Tracking of extended objects and group targets using random matrices},
  author={Feldmann, Michael and Franken, Dietrich and Koch, Wolfgang},
  journal={IEEE Transactions on Signal Processing},
  volume={59},
  number={4},
  pages={1409--1420},
  year={2011},
  publisher={IEEE}
}

@article{coraluppi2018multiple,
  title={Multiple-hypothesis tracking for targets producing multiple measurements},
  author={Coraluppi, Stefano P and Carthel, Craig A},
  journal={IEEE Transactions on Aerospace and Electronic Systems},
  volume={54},
  number={3},
  pages={1485--1498},
  year={2018},
  publisher={IEEE}
}

@article{cphdextended,
  title={An extended target {CPHD} filter and a gamma {G}aussian inverse {W}ishart implementation},
  author={Lundquist, Christian and Granstr{\"o}m, Karl and Orguner, Umut},
  journal={IEEE Journal of Selected Topics in Signal Processing},
  volume={7},
  number={3},
  pages={472--483},
  year={2013},
  publisher={IEEE}
}

@inproceedings{phdextended,
  title={Estimation and maintenance of measurement rates for multiple extended target tracking},
  author={Granstr{\"o}m, Karl and Orguner, Umut},
  booktitle={Proceedings of International Conference on Information Fusion},
  pages={2170--2176},
  year={2012},
  organization={IEEE}
}

@article{phdextended2,
  title={Extended target tracking using a {G}aussian-mixture {PHD} filter},
  author={Granstr{\"o}m, Karl and Lundquist, Christian and Orguner, Omut},
  journal={IEEE Transactions on Aerospace and Electronic Systems},
  volume={48},
  number={4},
  pages={3268--3286},
  year={2012},
  publisher={IEEE}
}

@article{lmbextended,
  title={Multiple extended target tracking with labeled random finite sets},
  author={Beard, Michael and Reuter, Stephan and Granstr{\"o}m, Karl and Vo, Ba-Tuong and Vo, Ba-Ngu and Scheel, Alexander},
  journal={IEEE Transactions on Signal Processing},
  volume={64},
  number={7},
  pages={1638--1653},
  year={2016},
  publisher={IEEE}
}

@article{pmbmextended2,
  title={Poisson multi-{B}ernoulli mixture conjugate prior for multiple extended target filtering},
  author={Granstr{\"o}m, Karl and Fatemi, Maryam and Svensson, Lennart},
  journal={IEEE Transactions on Aerospace and Electronic Systems},
  volume={56},
  number={1},
  pages={208--225},
  year={2020},
  publisher={IEEE}
}

@article{soextended,
  title={Likelihood-based data association for extended object tracking using sampling methods},
  author={Granstr{\"o}m, Karl and Svensson, Lennart and Reuter, Stephan and Xia, Yuxuan and Fatemi, Maryam},
  journal={IEEE Transactions on Intelligent Vehicles},
  volume={3},
  number={1},
  pages={30--45},
  year={2018},
  publisher={IEEE}
}

@inproceedings{xia2019extended,
  title={Extended target {P}oisson multi-{B}ernoulli mixture trackers based on sets of trajectories},
  author={Xia, Yuxuan and Granstr{\"o}m, Karl and Svensson, Lennart and Garc{\'i}a-Fern{\'a}ndez, {\'A}ngel F and Williams, Jason L},
  booktitle={22th International Conference on Information Fusion},
  pages={1--8},
  year={2019},
  organization={IEEE}
}

@inproceedings{gospa,
  title={Generalized optimal sub-pattern assignment metric},
  author={Rahmathullah, Abu Sajana and Garc{\'i}a-Fern{\'a}ndez, {\'A}ngel F and Svensson, Lennart},
  booktitle={Proceedings of International Conference on Information Fusion},
  pages={1--8},
  year={2017},
  organization={IEEE}
}

@article{pmbmpoint,
  title={Marginal multi-{B}ernoulli filters: {RFS} derivation of {MHT}, {JIPDA}, and association-based member},
  author={Williams, Jason L},
  journal={IEEE Transactions on Aerospace and Electronic Systems},
  volume={51},
  number={3},
  pages={1664--1687},
  year={2015},
  publisher={IEEE}
}

@article{pmbmpoint2,
  title={Poisson multi-{B}ernoulli mixture filter: direct derivation and implementation},
  author={Garc{\'i}a-Fern{\'a}ndez, {\'A}ngel F and Williams, Jason L and Granstr{\"o}m, Karl and Svensson, Lennart},
  journal={IEEE Transactions on Aerospace and Electronic Systems},
  volume={54},
  number={4},
  pages={1883--1901},
  year={2018},
  publisher={IEEE}
}

@inproceedings{gammareduction,
  title={On the reduction of {G}aussian inverse {W}ishart mixtures},
  author={Granstr{\"o}m, Karl and Orguner, Umut},
  booktitle={International Conference on Information Fusion},
  pages={2162--2169},
  year={2012},
  organization={IEEE}
}

@inproceedings{gwd,
  title={Metrics for performance evaluation of elliptic extended object tracking methods},
  author={Yang, Shishan and Baum, Marcus and Granstr{\"o}m, Karl},
  booktitle={International Conference on Multisensor Fusion and Integration for Intelligent Systems},
  pages={523--528},
  year={2016},
  organization={IEEE}
}

@inproceedings{ester1996density,
  title={A density-based algorithm for discovering clusters in large spatial databases with noise.},
  author={Ester, Martin and Kriegel, Hans-Peter and Sander, J{\"o}rg and Xu, Xiaowei and others},
  booktitle={Kdd},
  volume={96},
  number={34},
  pages={226--231},
  year={1996}
}

@article{vivone2016joint,
  title={Joint probabilistic data association tracker for extended target tracking applied to {X}-band marine radar data},
  author={Vivone, Gemine and Braca, Paolo},
  journal={IEEE Journal of Oceanic Engineering},
  volume={41},
  number={4},
  pages={1007--1019},
  year={2016},
  publisher={IEEE}
}

@ARTICLE{garcia2020trajectory,
  author={Garc{\'i}a-Fern{\'a}ndez, {\'A}ngel F and Svensson, Lennart and Williams, Jason L and Xia, Yuxuan and Granstr{\"o}m, Karl},
  journal={IEEE Transactions on Signal Processing}, 
  title={Trajectory {P}oisson Multi-{B}ernoulli Filters}, 
  year={2020},
  volume={68},
  number={},
  pages={4933-4945},
  doi={10.1109/TSP.2020.3017046}
}

@article{fatemi2017poisson,
  title={Poisson multi-{B}ernoulli mapping using {G}ibbs sampling},
  author={Fatemi, Maryam and Granstr{\"o}m, Karl and Svensson, Lennart and Ruiz, Francisco JR and Hammarstrand, Lars},
  journal={IEEE Transactions on Signal Processing},
  volume={65},
  number={11},
  pages={2814--2827},
  year={2017},
  publisher={IEEE}
}

@article{crouse2016implementing,
  title={On implementing 2{D} rectangular assignment algorithms},
  author={Crouse, David F},
  journal={IEEE Transactions on Aerospace and Electronic Systems},
  volume={52},
  number={4},
  pages={1679--1696},
  year={2016},
  publisher={IEEE}
}

@article{fortmann1983sonar,
  title={Sonar tracking of multiple targets using joint probabilistic data association},
  author={Fortmann, Thomas and Bar-Shalom, Yaakov and Scheffe, Molly},
  journal={IEEE journal of Oceanic Engineering},
  volume={8},
  number={3},
  pages={173--184},
  year={1983},
  publisher={IEEE}
}

@book{streit2021analytic,
  title={Analytic combinatorics in multiple object tracking},
  author={Streit, Roy L and Angle, R Blair and Efe, Murat},
  year={2021},
  publisher={Springer}
}

@article{garcia2021poisson,
  title={A {P}oisson multi-{B}ernoulli mixture filter for coexisting point and extended targets},
  author={Garc{\'i}a-Fern{\'a}ndez, {\'A}ngel F and Williams, Jason L and Svensson, Lennart and Xia, Yuxuan},
  journal={IEEE Transactions on Signal Processing},
  volume={69},
  pages={2600--2610},
  year={2021},
  publisher={IEEE}
}

@article{florian2021scalable,
  author={Meyer, Florian and Williams, Jason},
  journal={IEEE Transactions on Signal Processing}, 
  title={Scalable Detection and Tracking of Geometric Extended Objects}, 
  year={2021},
  volume={69},
  number={6283--6298},
  pages={}
}

@article{li2024multisensor,
  title={Multisensor multiple extended objects tracking based on the message passing},
  author={Li, Yuansheng and Shen, Tao and Gao, Lin},
  journal={IEEE Sensors Journal},
  volume={24},
  number={10},
  pages={16510--16528},
  year={2024},
  publisher={IEEE}
}

@article{guo2025gaussian,
  title={Gaussian Belief Propagation based Multi-View Multi-Extended Target Tracking with Occlusion},
  author={Guo, Yunfei and Zhang, Hao and Lin, Boting and Su, Hua and Chen, Yun},
  journal={IEEE Sensors Journal},
  year={2025},
  publisher={IEEE}
}

@article{xia2021poisson,
  title={Poisson Multi-{B}ernoulli Approximations for Multiple Extended Object Filtering},
  author={Xia, Yuxuan and Granstr{\"o}m, Karl and Svensson, Lennart and Fatemi, Maryam and Garc{\'i}a-Fern{\'a}ndez, {\'A}ngel F and Williams, Jason L},
  journal={IEEE Transactions on Aerospace and Electronic Systems},
  year={2022},
  volume={58},
  number={2},
  pages={890--906}
}

@article{granstrom2022tutorial,
  title={A Tutorial on Multiple Extended Object Tracking},
  author={Granstr{\"o}m, Karl and Baum, Marcus},
  year={2022},
  journal={TechRxiv}
}

@article{li2023adaptive,
  title={An adaptive and scalable multi-object tracker based on the non-homogeneous {P}oisson process},
  author={Li, Qing and Gan, Runze and Liang, Jiaming and Godsill, Simon J},
  journal={IEEE Transactions on Signal Processing},
  volume={71},
  pages={105--120},
  year={2023},
  publisher={IEEE}
}

@article{xia2023trajectory,
  title={Trajectory {PMB} filters for extended object tracking using belief propagation},
  author={Xia, Yuxuan and Garc{\'i}a-Fern{\'a}ndez, {\'A}ngel F and Meyer, Florian and Williams, Jason L and Granstr{\"o}m, Karl and Svensson, Lennart},
  journal={IEEE Transactions on Aerospace and Electronic Systems},
  year={2023},
  volume={59},
  number={6},
  pages={9312-9331},
  publisher={IEEE}
}

@inproceedings{xia2023efficient,
  title={An Efficient Implementation of the Extended Object Trajectory {PMB} Filter Using Blocked {G}ibbs Sampling},
  author={Xia, Yuxuan and Garc{\'i}a-Fern{\'a}ndez, {\'A}ngel F and Svensson, Lennart},
  booktitle={26th International Conference on Information Fusion (FUSION)},
  pages={1--8},
  year={2023},
  organization={IEEE}
}

@book{koller2009probabilistic,
  title={Probabilistic graphical models: principles and techniques},
  author={Koller, Daphne and Friedman, Nir},
  year={2009},
  publisher={MIT press}
}

@book{gelman1995bayesian,
  title={Bayesian data analysis},
  author={Gelman, Andrew and Carlin, John B and Stern, Hal S and Rubin, Donald B},
  year={1995},
  publisher={Chapman and Hall/CRC}
}

@article{ma2026closed,
  title={Closed-Form Message Passing Algorithms for Tracking Extended Targets},
  author={Ma, Weizhen and Jing, Zhongliang and Dong, Peng and Leung, Henry},
  journal={IEEE Transactions on Aerospace and Electronic Systems},
  year={2026},
  publisher={IEEE}
}

@article{ma2024max,
  title={Max Sum Based Data Associations for Tracking Point and Extended Targets},
  author={Ma, Weizhen and Jing, Zhongliang and Dong, Peng and Leung, Henry},
  journal={IEEE Transactions on Aerospace and Electronic Systems},
  year={2024},
  publisher={IEEE}
}

@article{cheng2025variational,
  title={Variational {B}ayesian Inference for Multiple Extended Targets or Unresolved Group Targets Tracking},
  author={Cheng, Yuanhao and Cao, Yunhe and Yeo, Tat-Soon and Zhang, Yulin and Fu, Jie},
  journal={IET Radar, Sonar \& Navigation},
  volume={19},
  number={1},
  pages={e70098},
  year={2025},
  publisher={Wiley Online Library}
}

@article{gan2024variational,
  title={Variational tracking and redetection for closely-spaced objects in heavy clutter},
  author={Gan, Runze and Li, Qing and Godsill, Simon J},
  journal={IEEE Transactions on Aerospace and Electronic Systems},
  volume={60},
  number={4},
  pages={5286--5311},
  year={2024},
  publisher={IEEE}
}

@inproceedings{li2023scalable,
  title={A scalable {R}ao-{B}lackwellised sequential {MCMC} sampler for joint detection and tracking in clutter},
  author={Li, Qing and Gan, Runze and Godsill, Simon},
  booktitle={26th International Conference on Information Fusion (FUSION)},
  pages={1--8},
  year={2023},
  organization={IEEE}
}

@inproceedings{gan2025pivot,
  title={PiVoT: {P}oisson measurements-based variational multi-object detection and tracking},
  author={Gan, Runze and Li, Qing and Hopgood, James R and Davies, Mike E and Godsill, Simon},
  booktitle={28th International Conference on Information Fusion (FUSION)},
  pages={1--8},
  year={2025},
  organization={IEEE}
}

@article{baerveldt2026combining,
  title={Combining Occupancy Grid Mapping and Extended Object Tracking With the {P}oisson Multi-{B}ernoulli Mixture Filter},
  author={Baerveldt, Martin and Svensson, Lennart and Brekke, Edmund F{\o}rland},
  journal={IEEE Journal of Oceanic Engineering},
  year={2026},
  publisher={IEEE}
}

@article{baerveldt2024multiple,
  title={A Multiple Extended Object Tracker with the {G}aussian Process Model Utilizing Negative Information},
  author={Baerveldt, Martin and L\'{o}pez, Michael Ernesto and Brekke, Edmund F{\o}rland},
  journal={Journal of Advances in Information Fusion},
  volume={19},
  number={1},
  pages={88--108},
  year={2024},
  publisher={IEEE}
}

@inproceedings{liu2024framework,
  title={Which framework is suitable for online 3{D} multi-object tracking for autonomous driving with automotive 4{D} imaging radar?},
  author={Liu, Jianan and Ding, Guanhua and Xia, Yuxuan and Sun, Jinping and Huang, Tao and Xie, Lihua and Zhu, Bing},
  booktitle={IEEE Intelligent Vehicles Symposium (IV)},
  pages={1258--1265},
  year={2024},
  organization={IEEE}
}

@article{ge20205g,
  title={5{G} {SLAM} using the clustering and assignment approach with diffuse multipath},
  author={Ge, Yu and Wen, Fuxi and Kim, Hyowon and Zhu, Meifang and Jiang, Fan and Kim, Sunwoo and Svensson, Lennart and Wymeersch, Henk},
  journal={Sensors},
  volume={20},
  number={16},
  pages={4656},
  year={2020},
  publisher={MDPI}
}

@inproceedings{ding2024lidar,
  title={{LiDAR} point cloud-based multiple vehicle tracking with probabilistic measurement-region association},
  author={Ding, Guanhua and Liu, Jianan and Xia, Yuxuan and Huang, Tao and Zhu, Bing and Sun, Jinping},
  booktitle={27th International Conference on Information Fusion (FUSION)},
  pages={1--8},
  year={2024},
  organization={IEEE}
}

@inproceedings{xia2024bayesian,
  title={{B}ayesian simultaneous localization and multi-lane tracking using onboard sensors and a {SD} map},
  author={Xia, Yuxuan and Stenborg, Erik and Fu, Junsheng and Hendeby, Gustaf},
  booktitle={27th International Conference on Information Fusion (FUSION)},
  pages={1--8},
  year={2024},
  organization={IEEE}
}

@article{deng20243,
  title={3-{D} multiple extended object tracking by fusing roadside radar and camera sensors},
  author={Deng, Jiayin and Hu, Zhiqun and Lu, Zhaoming and Wen, Xiangming},
  journal={IEEE Sensors Journal},
  volume={25},
  number={1},
  pages={1885--1899},
  year={2024},
  publisher={IEEE}
}

@article{yang2025three,
  title={Three-dimensional Multiple Extended Targets Tracking under occlusion using Variational {G}aussian Processes},
  author={Yang, Dongsheng and Guo, Yunfei and Li, Xiaotian and Chen, Yun and Shentu, Han},
  journal={IEEE Transactions on Aerospace and Electronic Systems},
  volume={61},
  number={5},
  pages={12951-12969},
  year={2025},
  publisher={IEEE}
}

@book{bar2011tracking,
  title={Tracking and data fusion},
  author={Bar-Shalom, Yaakov and Willett, Peter K and Tian, Xin},
  volume={11},
  year={2011},
  publisher={YBS publishing Storrs, CT, USA:}
}

@article{zhu2021detection,
  title={Detection and tracking meet drones challenge},
  author={Zhu, Pengfei and Wen, Longyin and Du, Dawei and Bian, Xiao and Fan, Heng and Hu, Qinghua and Ling, Haibin},
  journal={IEEE Transactions on Pattern Analysis and Machine Intelligence},
  volume={44},
  number={11},
  pages={7380--7399},
  year={2021},
  publisher={IEEE}
}

@inproceedings{cioppa2022soccernet,
  title={Soccernet-tracking: {M}ultiple object tracking dataset and benchmark in soccer videos},
  author={Cioppa, Anthony and Giancola, Silvio and Deliege, Adrien and Kang, Le and Zhou, Xin and Cheng, Zhiyu and Ghanem, Bernard and Van Droogenbroeck, Marc},
  booktitle={Proceedings of the IEEE/CVF Conference on Computer Vision and Pattern Recognition},
  pages={3491--3502},
  year={2022}
}

@article{huang2025vehicle,
  title={Vehicle-to-everything cooperative perception for autonomous driving},
  author={Huang, Tao and Liu, Jianan and Zhou, Xi and Nguyen, Dinh C and Azghadi, Mostafa Rahimi and Xia, Yuxuan and Han, Qing-Long and Sun, Sumei},
  journal={Proceedings of the IEEE},
  year={2025},
  volume={113},
  number={5},
  pages={443-477},
  publisher={IEEE}
}

@article{garcia2023poisson,
  title={Poisson multi-{B}ernoulli mixture filter with general target-generated measurements and arbitrary clutter},
  author={Garc{\'\i}a-Fern{\'a}ndez, {\'A}ngel F and Xia, Yuxuan and Svensson, Lennart},
  journal={IEEE Transactions on Signal Processing},
  volume={71},
  pages={1895--1906},
  year={2023},
  publisher={IEEE}
}

@article{granstrom2024poisson,
  title={Poisson multi-{B}ernoulli mixtures for sets of trajectories},
  author={Granstr{\"o}m, Karl and Svensson, Lennart and Xia, Yuxuan and Williams, Jason and Garc{\'\i}a-Fern{\'a}ndez, {\'A}ngel F},
  journal={IEEE Transactions on Aerospace and Electronic Systems},
  volume={61},
  number={2},
  pages={5178--5194},
  year={2024},
  publisher={IEEE}
}

@article{van2008partially,
  title={Partially collapsed {G}ibbs samplers: {T}heory and methods},
  author={Van Dyk, David A and Park, Taeyoung},
  journal={Journal of the American Statistical Association},
  volume={103},
  number={482},
  pages={790--796},
  year={2008},
  publisher={Taylor \& Francis}
}

@inproceedings{xie2026bfmap,
  title     = {{B$^2$F}-Map: {C}rowd-sourced Mapping with {B}ayesian {B}-spline Fusion},
  author    = {Xie, Yiping and Xia, Yuxuan and Stenborg, Erik and Fu, Junsheng and Beauvisage, Axel and Garcia, Gabriel E. and Wu, Tianyu and Hendeby, Gustaf},
  booktitle = {Proceedings of the IEEE International Conference on Robotics and Automation (ICRA)},
  year      = {2026},
  note      = {Accepted. arXiv:2603.01673},
}

@article{tierney1994markov,
  title={Markov chains for exploring posterior distributions},
  author={Tierney, Luke},
  journal={the Annals of Statistics},
  pages={1701--1728},
  year={1994},
  publisher={JSTOR}
}

\newpage
\appendices
\textbf{\huge \center Supplementary Materials}

\section{Derivation of Proposition~\ref{prop:approx_a1}}
\label{app:approx_a1}

We first recall the KL projection onto a product-form family. Let $y_i=(u^i,x_k^i)$ and $y_{1:n_{k|k-1}}=\{(u^i,x_k^i)\}_{i=1}^{n_{k|k-1}}$. Also let $p_a(y_{1:n_{k|k-1}})$ denote the distribution corresponding to the factor $\widetilde{f}^a_{k|k}(\{(u^i,x_k^i)\}_{i=1}^{n_{k|k-1}})$ in \eqref{eq_marginal_global_hypothesis}. Consider the product-form approximation
\begin{equation}
  q_a\left(y_{1:n_{k|k-1}}\right)=\prod_{i=1}^{n_{k|k-1}}q^{i,a^i}(y_i).
\end{equation}
The KL divergence can be written as
\begin{align}
  &D\left(p_a\left(y_{1:n_{k|k-1}}\right) \parallel \prod_{i=1}^{n_{k|k-1}}q^{i,a^i}(y_i)\right) \nonumber\\
  &= \mathbb{E}_{p_a}\left[\log p_a\left(y_{1:n_{k|k-1}}\right)\right]-\sum_{i=1}^{n_{k|k-1}}\mathbb{E}_{p_a}\left[\log q^{i,a^i}(y_i)\right] \nonumber\\
  &= C_a+\sum_{i=1}^{n_{k|k-1}}D\left(p_a^i(y_i) \parallel q^{i,a^i}(y_i)\right),
  \label{eq_appendix_a1_kl}
\end{align}
where $p_a^i(\cdot)$ is the $i$-th marginal of the joint $p_a(\cdot)$, and $C_a$ is a constant independent of the factors $q^{i,a^i}(\cdot)$. Therefore, the minimum is attained, for $i\in\{1,\dots,n_{k|k-1}\}$, when
\begin{equation}
  q^{i,a^i}(y_i)=p_a^i(y_i).
\end{equation}

For the factorization in \eqref{eq_marginal_global_hypothesis}, the required component marginal is obtained by summing over the corresponding object existence variable:
\begin{equation}
  q^{i,a^i}\left(u^i,x_k^i\right) \propto \sum_{e_k^i\in\{0,1\}}\underline{f}^{i,a^i}_{k|k-1}\left(u^i,x_k^i,e_k^i\right)p\left(e_k^i\right),
  \label{eq_appendix_a1_marginal}
\end{equation}
where $p(e_k^i=1)=r^{i,\widetilde{a}^i}_{k|k}$ and $p(e_k^i=0)=1-r^{i,\widetilde{a}^i}_{k|k}$. For $e_k^i=1$, substitution of \eqref{eq_existing_prior} gives the contribution
\begin{equation}
  \delta_i\left[u^i\right]r^{i,a^i}_{k|k-1}e^{-\gamma_k\left(x_k^i\right)}f^{i,a^i}_{k|k-1}\left(x_k^i\right)r^{i,\widetilde{a}^i}_{k|k}.
\end{equation}
For $e_k^i=0$, the corresponding contribution is
\begin{equation}
  \left(1-r^{i,a^i}_{k|k-1}\right)\left(1-r^{i,\widetilde{a}^i}_{k|k}\right).
\end{equation}
Adding these two terms yields \eqref{eq_predicted_global_hypo_marginal_existence}. Under Approximation A1, we substitute
\begin{equation}
  \widetilde{f}^a_{k|k}\left(\left\{\left(u^i,x_k^i\right)\right\}_{i=1}^{n_{k|k-1}}\right) \approx\prod_{i=1}^{n_{k|k-1}}q^{i,a^i}\left(u^i,x_k^i\right)
\end{equation}
into \eqref{eq_marginal_global_hypothesis}. The proportionality in \eqref{eq_marginal_global_hypothesis} then directly gives the approximate conditional distribution in \eqref{eq_collapsed_sample_a}.

\section{Derivation of Proposition~\ref{prop:approx_a2}}
\label{app:approx_a2}

Let
\begin{equation}
  p_{\beta}(\beta_k)={f}_{k|k}\left(\beta_k | \left\{\left(u^i,x_k^i\right)\right\}_{i=1}^{n_{k|k}}\right)
\end{equation}
denote the exact collapsed conditional distribution in \eqref{eq_marginal_data_association}. For the product-form family in Approximation A2, the KL divergence satisfies
\begin{align}
  &D\left(p_{\beta}(\beta_k)\parallel \prod_{j=1}^{m_k}q^j(\beta_k^j)\right) \nonumber\\
  &=C_{\beta}+\sum_{j=1}^{m_k}D\left(p_{\beta}^j(\beta_k^j)\parallel q^j(\beta_k^j)\right),
  \label{eq_appendix_a2_kl}
\end{align}
where $p_{\beta}^j(\cdot)$ is the marginal distribution of the $j$-th association variable, and $C_{\beta}$ does not depend on the factors $q^j(\cdot)$. Hence, the minimum is attained at
\begin{equation}
  q^j(\beta_k^j)=p_{\beta}^j\left(\beta_k^j\right).
\end{equation}

The $j$-th marginal can be expressed by summing over the object existence variables:
\begin{align}
  &q^j\left(\beta_k^j | \left\{\left(u^i,x_k^i\right)\right\}_{i=1}^{n_{k|k}}\right) \nonumber\\
  &=\sum_{e_k^{1:n_{k|k}}}{f}_{k|k}\left(\beta_k^j | \left\{\left(u^i,x_k^i,e_k^i\right)\right\}_{i=1}^{n_{k|k}}\right)\prod_{i=1}^{n_{k|k}}p\left(e_k^i\right).
  \label{eq_appendix_a2_marginal}
\end{align}

For a candidate assignment $\beta_k^j=i$, the exact conditional distribution in \eqref{eq_sample_beta1} and \eqref{eq_sample_beta2} depends only on the potential object assigned to the measurement and on the new Bernoulli component $i^\prime=n_{k|k-1}+j$ initiated by $z_k^j$. Therefore, the sums over all other existence variables in \eqref{eq_appendix_a2_marginal} are equal to one, and the marginal reduces to
\begin{align}
  &q^j\left(\beta_k^j=i | \left(u^i,x_k^i\right),\left(u^{i^\prime},x_k^{i^\prime}\right)\right) \nonumber\\
  &=\sum_{e_k^i,e_k^{i^\prime}}{f}_{k|k}\left(\beta_k^j=i | \left(u^i,x_k^i,e_k^i\right),\left(u^{i^\prime},x_k^{i^\prime},e_k^{i^\prime}\right)\right) \nonumber\\
  &~~~~~~~~~\times p\left(e_k^i\right)p\left(e_k^{i^\prime}\right).
  \label{eq_appendix_a2_reduced}
\end{align}

If $i$ is an existing Bernoulli component, assigning $z_k^j$ to $i$ requires component $i$ to exist and the potential new component $i^\prime$ to be absent. The corresponding existence probability is
\begin{equation}
  r^{i,\widetilde{a}^i}_{k|k}\left(1-r^{i^\prime,\widetilde{a}^{i^\prime}}_{k|k}\right).
\end{equation}
Multiplying this probability by the measurement likelihood factor in \eqref{eq_sample_beta1} gives \eqref{eq_collapsed_sample_beta1}.

If $i$ is a new Bernoulli component with $j<i-n_{k|k-1}$, measurement $z_k^j$ is assigned to a new component initiated by a later-indexed measurement. In this case, component $i$ must exist and component $i^\prime$ must be absent, which yields the first case of \eqref{eq_collapsed_sample_beta2}. If $j=i-n_{k|k-1}$, then $i=i^\prime$, and the measurement is assigned to the $j$-th new Bernoulli component. The two possible existence states give
\begin{equation}
  \delta_i\left[u^i\right]\gamma_k\left(x_k^i\right)\ell_k\left(z_k^j \mid x_k^i\right)r^{i,\widetilde{a}^i}_{k|k} +\lambda_k^C\left(z_k^j\right)\left(1-r^{i,\widetilde{a}^i}_{k|k}\right),
\end{equation}
corresponding respectively to a newly detected object and clutter. This gives the second case of \eqref{eq_collapsed_sample_beta2}. Assignments with $j>i-n_{k|k-1}$ are outside the support of $\beta_k^j$ and therefore have probability zero.

\section{Computational Complexity}
\label{app:complexity}

We analyze the computational complexity of the sampling operations in Algorithm~\ref{alg_full_gibbs}. Evaluations and samples of fixed-dimensional single object factors are treated as operations with constant cost.

First, sampling the set of object states involves at most $n_{k|k}=n_{k|k-1}+m_k$ Bernoulli components and therefore requires $\mathcal{O}(n_{k|k})$ operations per iteration.

Sampling the previous global hypothesis requires evaluating the conditional probability in \eqref{eq_sample_a} or \eqref{eq_collapsed_sample_a} for every retained hypothesis in $\mathcal{A}_{k|k-1}$. Each candidate contains at most $n_{k|k-1}$ Bernoulli component factors, resulting in
\begin{equation}
    \mathcal{O}\left(
    |\mathcal{A}_{k|k-1}|n_{k|k-1}
    \right)
\end{equation}
operations per iteration.

For each measurement $z_k^j$, sampling $\beta_k^j$ requires evaluating and normalizing a categorical distribution over at most $n_{k|k}$ assignments. Thus, sampling all $m_k$ association variables requires
\begin{equation}
    \mathcal{O}(m_kn_{k|k})
    =
    \mathcal{O}\left(
    m_k(n_{k|k-1}+m_k)
    \right)
\end{equation}
total computational operations. Note that since the measurement association variables are conditionally independent, these operations can be performed in parallel. 

Combining the three sampling steps, the complexity of one complete Gibbs iteration is
\begin{align}
    \mathcal{O}\Big(
    &n_{k|k}
    +|\mathcal{A}_{k|k-1}|n_{k|k-1}
    +m_kn_{k|k}
    \Big)
    \nonumber\\
    &=
    \mathcal{O}\left(
    |\mathcal{A}_{k|k-1}|n_{k|k-1}
    +m_k(n_{k|k-1}+m_k)
    \right).
\end{align}
The total sampling complexity over $N_{\text{iter}}$ iterations is 
\begin{equation}
    \mathcal{O}\left(
    N_{\text{iter}}
    \left[
    |\mathcal{A}_{k|k-1}|n_{k|k-1}
    +m_k(n_{k|k-1}+m_k)
    \right]
    \right).
\end{equation}

The expressions above give a worst-case upper bound. In the track-oriented implementation, the measurement association and updated density of each local hypothesis are stored in hash tables. If the same local hypothesis is encountered in multiple sampling iterations, its updated density is retrieved rather than recomputed. Consequently, the expensive single object update is performed only once for each unique local hypothesis. Thus, the practical local hypothesis update cost scales with the number of unique local hypotheses encountered rather than with the total number of visits to those hypotheses.

Finally, the implementations generate at most $N_{\text{iter}}$ updated global hypotheses before duplicate removal, rather than explicitly enumerating the complete updated global hypothesis space, whose cardinality grows combinatorially with $m_k$. The number of predicted global hypotheses $|\mathcal{A}_{k|k-1}|$ is controlled by PMBM pruning.

\section{Proof of Proposition~\ref{prop:gibbs_convergence}}
\label{app:gibbs_convergence}

Let
\begin{equation}
    s=\left(\widetilde{\mathbf{x}}_k^d,a,\beta_k\right)
\end{equation}
denote the state of the Markov chain, and let $\mathsf{S}_k$ denote its feasible joint state space, which is 
\begin{align}
\mathsf{S}_k \triangleq
\Big\{
&(\widetilde{\mathbf{x}}_k^d,a,\beta_k):
\widetilde{\mathbf{x}}_k^d
\in
\mathcal{F}\left(
(\mathbb{U}_{k|k}\setminus\{0\})\times\mathcal{X}
\right),
a\in\mathcal{A}_{k|k-1},
\nonumber\\
&
\beta_k^j\in
\{1,\dots,n_{k|k-1}\}
\cup
\{n_{k|k-1}+j,\dots,n_{k|k}\},
\nonumber\\
&\forall j\in\{1,\dots,m_k\},
{f}_{k|k}^d
(\widetilde{\mathbf{x}}_k^d,a,\beta_k)>0
\Big\},
\end{align}
where the last condition ${f}_{k|k}^d
(\widetilde{\mathbf{x}}_k^d,a,\beta_k)>0$ enforces the feasibility constraints. The space $\mathsf{S}_k$ is further equipped with the sigma-algebra
$\mathcal{B}(\mathsf{S}_k)$ induced by the corresponding product space.

We further let $\Pi_k$ denote the probability measure on $\mathsf{S}_k$ induced by the joint posterior density
${f}_{k|k}^d(\widetilde{\mathbf{x}}_k^d,a,\beta_k)$. Also, let $K_{\mathbf{x}}$, $K_a$, and $K_{\beta}$ denote the Markov transition kernels of the full Gibbs sampler on $\mathsf{S}_k$ that update the set of detected object states \eqref{eq_set_detected_conditional}, the previous global hypothesis \eqref{eq_sample_a}, and the measurement association vector \eqref{eq_sample_beta}, respectively, while leaving the other blocks unchanged. 

Since each block of the full Gibbs sampler is drawn from its exact
conditional distribution, the corresponding block update kernel leaves
$\Pi_k$ invariant \cite[Sec.~2.2]{tierney1994markov}.
Thus, for any $B\in\mathcal{B}(\mathsf{S}_k)$,
\begin{align}
    \int_{\mathsf{S}_k}\Pi_k(\mathrm{d}s)K_{\mathbf{x}}(s,B)
    &=\Pi_k(B),\\
    \int_{\mathsf{S}_k}\Pi_k(\mathrm{d}s)K_a(s,B)
    &=\Pi_k(B),\\
    \int_{\mathsf{S}_k}\Pi_k(\mathrm{d}s)K_{\beta}(s,B)
    &=\Pi_k(B).
\end{align}
The transition kernel of one complete Gibbs iteration is
\begin{align}
    P(s,B)
    =
    \int_{\mathsf{S}_k}\int_{\mathsf{S}_k}
    &K_{\mathbf{x}}(s,\mathrm{d}s_1)
    K_a(s_1,\mathrm{d}s_2)
    K_{\beta}(s_2,B).
\end{align}
Since the three kernels share the same invariant distribution, their composition also leaves $\Pi_k$ invariant:
\begin{equation}
    \int_{\mathsf{S}_k}\Pi_k(\mathrm{d}s)P(s,B)
    =
    \Pi_k(B).
\end{equation}

According to the assumptions in Proposition~\ref{prop:gibbs_convergence}, the exact full conditionals are positive on their feasible supports. Hence, \eqref{eq_sample_a} allows every previous global hypothesis compatible with the set of detected object states to be selected. For an existing Bernoulli component, \eqref{eq_sample_beta1} permits any feasible measurement assignment when the object is present and the new Bernoulli component initiated by the same measurement is absent. In addition, \eqref{eq_set_detected_conditional} allows an existing Bernoulli component with no assigned measurements to be sampled as present under its missed detection hypothesis whenever this state is feasible, after which it can be associated with measurements. Existing Bernoulli components therefore impose no anchoring constraint on the association space. 

To prove $\Pi_k$-irreducibility
\cite[Sec. 3.1]{tierney1994markov}, it remains to show that the Markov
chain can move, through finitely many Gibbs updates with positive
probability, between any two feasible measurement-association
configurations involving newly detected objects. As shown by \eqref{eq_sample_beta2}, the $j$-th new Bernoulli component
is anchored to its initiating measurement $z_k^j$ whenever it exists, so
this measurement cannot be reassigned by the separate block updates unless
the Bernoulli component first becomes absent. The following argument shows that the positive-clutter condition in Proposition~\ref{prop:gibbs_convergence} provides precisely this transition path.

For the local hypothesis associated with measurement $z_k^j$ of the $j$-th new Bernoulli component, whose index is $i=n_{k|k-1}+j$, \eqref{eq_newBernoulli2} gives
\begin{equation}
    r_{k|k}^{i,a^i}
    =
    \frac{\ell_{k|k}^{i,\iota}}
    {\lambda_k^C(z_k^j)+\ell_{k|k}^{i,\iota}},
\end{equation}
which is strictly between zero and one since $\lambda_k^C(z_k^j)>0$. This new Bernoulli component can therefore become an empty set while $\beta_k^j=i$, in which case $z_k^j$ is interpreted as clutter. Once the component is absent, the measurement can be reassigned to another Bernoulli component with positive probability. Conversely, a measurement assigned elsewhere could first be assigned to the $j$-th new Bernoulli component created by that measurement while the component remains absent, corresponding to clutter. At the subsequent sampling step, the Bernoulli set can be sampled as a singleton, changing the interpretation of the measurement from clutter to a newly detected object.

We note that a local hypothesis associated with more than one measurement has existence probability one, but this does not prevent transitions to other associations. Its non-anchoring measurements can be reassigned individually until only the anchoring measurement remains. The sampler can then pass through the singleton clutter configuration with non-zero probability. Therefore, positive clutter intensity $\lambda_k^C(\cdot)$ connects the feasible association configurations. Together with positivity of the remaining conditional distributions, this establishes irreducibility on the feasible posterior support.

Moreover, the conditional distributions assign positive probability to leaving the previous global hypothesis and measurement association unchanged, and also allow the continuous object states to remain within a set of positive posterior measure. The chain can therefore return to the same positive measure region in consecutive iterations and is aperiodic. Since $P$ is $\Pi_k$-irreducible and aperiodic and has invariant
probability distribution $\Pi_k$, \cite[Thm.~1]{tierney1994markov}
implies that $\Pi_k$ is the unique stationary distribution and, for
$\Pi_k$-almost every feasible initial state $s$, the total variation
distance between the $t$-step transition distribution $P^t(s,\cdot)$
and $\Pi_k$ approaches zero:
\begin{equation}
    \lim_{t\rightarrow\infty}
    \sup_{B\in\mathcal{B}(\mathsf{S}_k)}
    \left|
    P^t(s,B)-\Pi_k(B)
    \right|
    =0.
\end{equation}
This states that, as the number of sampling iterations increases, the probability $P^t(s,B)$ assigned by the sampler after $t$ iteration converges uniformly over all measurable events $B$ to the posterior probability $\Pi_k(B)$.

\section{GGIW Implementation Details}
\label{app:ggiw_implementation}

For the dynamic model, we assume that the kinematic state $\xi_k$ follows a linear Gaussian transition model, and that the measurement rate $\gamma_k$ and the extent matrix $X_k$ evolve slowly over time. In particular,
\begin{equation}
  \xi_{k} = F\xi_{k-1} + w_k, \quad \gamma_{k} = \gamma_{k-1}, \quad X_{k} = X_{k-1},
  \label{eq_ggiw_motion_model}
\end{equation}
where $F$ is the transition matrix, and $w_k$ is a zero-mean Gaussian process noise with covariance matrix $Q$. We further assume a constant object survival probability $p^S$.

The GGIW implementation recursively computes the parameter vector
\begin{equation}
\zeta_{k|k^\prime} = \left\{\alpha_{k|k^\prime},\beta_{k|k^\prime},\bar{x}_{k|k^\prime},P_{k|k^\prime},v_{k|k^\prime},V_{k|k^\prime}\right\},
\end{equation}
with $k^\prime\in\{k-1,k\}$ over time. Using the motion model in \eqref{eq_ggiw_motion_model} and the GGIW parameters $\zeta_{k-1|k-1}$ at time step $k-1$, the predicted parameters $\zeta_{k|k-1}$ at time step $k$ can be obtained as \cite{pmbmextended2}
\begin{subequations}
  \label{eq_ggiw_motion}
  \begin{align}
    \alpha_{k|k-1} &= \alpha_{k-1|k-1}/\eta, \\
    \beta_{k|k-1} &= \beta_{k-1|k-1}/\eta, \\
    \bar{x}_{k|k-1} &= F\bar{x}_{k-1|k-1}, \\
    P_{k|k-1} &= FP_{k-1|k-1}F^T + Q, \\
    v_{k|k-1} &= d + 1 + e^{-T_s/\tau}(v_{k-1|k-1}-d-1), \\
    V_{k|k-1} &= e^{-T_s/\tau}V_{k-1|k-1}.
  \end{align}
\end{subequations}
Here, $\eta>1$ is a forgetting factor governing the evolution of the measurement rate, $T_s$ is the sampling interval, and $\tau>0$ is a time constant that regulates the rate of change of the object extent.

Given the PPP measurement model \eqref{eq_ppp_meas}, the measurement likelihood \eqref{eq_single_measurement_likelihood}, and measurements $\mathbf{w}$, the updated GGIW parameters $\zeta_{k|k}$ can be computed as in \cite{pmbmextended2}:
\begin{subequations}\label{eq_ggiw_update}
  \begin{align}
    \alpha_{k|k} &= \alpha_{k|k-1} + |\mathbf{w}|, \\
    \beta_{k|k} &= \beta_{k|k-1} + 1, \\
    \bar{x}_{k|k} &= \bar{x}_{k|k-1} + K\varepsilon, \\
    P_{k|k} &= P_{k|k-1} - KHP_{k|k-1}, \\
    v_{k|k} &= v_{k|k-1} + |\mathbf{w}|, \\
    V_{k|k} &= V_{k|k-1} + N + Z.
  \end{align}
\end{subequations}
The required quantities are
\begin{subequations}\label{eq_ggiw_update_paras}
  \begin{align}
    \bar{{z}} &= \textstyle \frac{1}{|\mathbf{w}|}\sum_{z\in \mathbf{w}} z, \\
    Z &= \textstyle \sum_{z\in \mathbf{w}}(z - \bar{{z}})(z - \bar{{z}})^T, \\
    \hat{X} &= V_{k|k-1}/(v_{k|k-1} - d - 1), \\
    \varepsilon &= \bar{{z}} - H\bar{x}_{k|k-1}, \\
    S &= H P_{k|k-1} H^T + \hat{X}/|\mathbf{w}|, \\
    K &= P_{k|k-1} H^T S^{-1}, \\
    N &= \hat{X}^{1/2} S^{-1/2} \varepsilon \varepsilon^T \left(S^{-1/2}\right)^T \left(\hat{X}^{1/2}\right)^T.
  \end{align}
\end{subequations}
The predicted likelihood is
\begin{align}
  \ell_{k|k} &= \left(\pi^{|\mathbf{w}|}|\mathbf{w}|\right)^{-\frac{d}{2}}\nonumber\\
  &~~~\times \frac{|V_{k|k-1}|^{\frac{v_{k|k-1}}{2}}\Gamma_d(v_{k|k})\left|\hat{X}\right|^{\frac{1}{2}}\Gamma(\alpha_{k|k})\beta_{k|k-1}^{\alpha_{k|k-1}}}{|V_{k|k}|^{\frac{v_{k|k}}{2}}\Gamma_d(v_{k|k-1})|S|^{\frac{1}{2}}\Gamma(\alpha_{k|k-1})\beta_{k|k}^{\alpha_{k|k}}},\label{eq_predicted_likelihood}
\end{align}
where $\Gamma(\cdot)$ and $\Gamma_d(\cdot)$ denote the scalar Gamma function and its multivariate counterpart, respectively. In the case of misdetection, the predicted likelihood reduces to
\begin{equation}
  \ell_{k|k} = \left(\frac{\beta_{k|k-1}}{\beta_{k|k-1}+1}\right)^{\alpha_{k|k-1}}. \label{eq_predicted_missed_likelihood}
\end{equation}
The predicted likelihoods in \eqref{eq_predicted_likelihood} and \eqref{eq_predicted_missed_likelihood} are incorporated into the PMBM update described in Section II-C (cf. \eqref{eq_likelihood_missed_ppp}, \eqref{eq_likelihood_missed}, \eqref{eq_likelihood_Bernoulli_update}, and \eqref{eq_detection_likelihood_undetected}), and are used to evaluate the updated global hypothesis weights $w_{k|k}^a$.

The PPP intensity of undetected objects is represented as a GGIW mixture,
\begin{equation}
  \lambda_{k|k^\prime}(x_k) = \sum_{l=1}^{N_{k|k^\prime}^u} w_{k|k^\prime}^{u,l}\mathcal{GGIW}\left(x_k ; \zeta^{u,l}_{k|k^\prime}\right), \label{eq_ggiw_ppp}
\end{equation}
with $N_{k|k^\prime}^u$ mixture components, where $w_{k|k^\prime}^{u,l}>0$ is the weight of the $l$-th component. To perform measurement update of the PPP intensity \eqref{eq_ggiw_ppp}, we first apply the GGIW update in \eqref{eq_ggiw_update} to each mixture component separately, and then merge all the resulting components into a single one
using, e.g., the density-reduction methods in \cite{phdextended,gammareduction}.

\end{document}